\documentclass[11pt,a4paper]{article}

\input{preamble.tex}

\newcommand{\p}{\mathrm{P}}
\newcommand{\g}{\mathsf{g}} 
\newcommand{\s}[1]{{\scriptscriptstyle #1}}
\newcommand{\V}{\sfV_{d-1}}
\newcommand{\vlm}{{\ell\vec{m}}}
 
\newcommand{\vm}{{\vec{m}}} 

\begin{document}

\begin{center}
	{\Large\textsc{Higher-form anomalies and state-operator correspondence beyond conformal invariance}} \\
	\bigskip
	Stathis Vitouladitis \\
	\bigskip
	\footnotesize{
		\textrm{Physique Théorique et Mathématique, \\ Université Libre de Bruxelles \& International Solvay Institutes,\\ CP 231, 1050 Brussels, Belgium} \\
	}
	\bigskip
	\href{mailto:stathis.vitouladitis@ulb.be}{\small \sf stathis.vitouladitis@ulb.be}
\end{center}

\vspace{3em}

\begin{abstract}
	\noindent We establish a state-operator correspondence for a class of non-conformal quantum field theories with continuous higher-form symmetries and a mixed anomaly. Such systems can always be realised as a relativistic superfluid. The symmetry structure induces an infinite tower of conserved charges, which we construct explicitly. These charges satisfy an abelian current algebra with a central extension, generalising the familiar Kac--Moody algebras to higher dimensions. States and operators are organised into representations of this algebra, enabling a direct correspondence. We demonstrate the correspondence explicitly in free examples by performing the Euclidean path integral on a \(d\)-dimensional ball, with local operators inserted in the origin, and matching to energy eigenstates on \(S^{d-1}\) obtained by canonical quantisation. Interestingly, in the absence of conformal invariance, the empty path integral prepares a squeezed vacuum rather than the true ground state.
\end{abstract}

\pagebreak
{\hypersetup{linkcolor=black}
	\tableofcontents
	\vspace{10pt}
}

\section{Introduction}\label{sec:intro}

One of the most striking and powerful features of conformal field theory (CFT) is the \emph{state-operator correspondence}. At first glance, it seems almost paradoxical: it  asserts that every quantum state defined on a spatial sphere can be associated with a local operator inserted at a single point in spacetime. This is surprising because states are inherently non-local objects, defined on an entire Cauchy surface, while local operators are the most ultra-localised observables the theory permits.

Nonetheless, conformal invariance ensures that this is the case. In short, it is a consequence of a conformal transformation from the Euclidean cylinder, \(\R\times\S^{d-1}\), to the plane, which maps the infinite Euclidean past to a single point (which we shall call the \emph{origin}) on the plane. Under this map the Hamiltonian, generating time translations on the cylinder, gets replaced by the dilatation operator; that is \(\cD=r\,\pd_r\) in spherical coordinates. Hence, preparing energy eigenstates amounts to imposing a local condition at the origin, that transforms appropriately under rescalings. This is precisely a local operator.\footnote{The story is of course much more concrete than the cartoon we presented here. See for instance \cite{Polchinski:1998rq} for a good account.}

The consequences of the state-operator correspondence are nothing short of dramatic. To name only a few: for CFTs themselves, in two dimensions, it allows the theory to be consistently defined on arbitrary Riemann surfaces \cite{Moore:1988qv}. This is, in turn, a crucial ingredient in the internal consistency of string theory. In the same spirit, the correspondence is indispensable for the conformal bootstrap \cite{Rattazzi:2008pe,Poland:2018epd,Hartman:2022zik,Poland:2022qrs}, guiding the search for extracting consistent CFTs and their spectrum of local operators and scaling dimensions. In critical phenomena, it provides a very effective tool to study entanglement entropy, particularly via the replica trick \cite{Calabrese:2009qy}. There are also deep implications for gravity, most notably via holography and the AdS/CFT correspondence. Perhaps the most striking of these is that black hole microstates are probed by “heavy operators” in the dual CFT \cite{Maldacena:1997re,Witten:1998qj,Aharony:1999ti}.\footnote{Honourable mentions of recent promising approaches where the state-operator correspondence plays a crucial role include the large-charge expansion \cite{Hellerman:2015nra,Monin:2016jmo}, as well as fuzzy-sphere methods \cite{Zhu:2022gjc}.}

Sadly, the arguments outlined above do not extend to generic non-conformal quantum field theories (QFTs). One direction of the correspondence remains trivially valid: acting with a local operator on the vacuum produces a new state. More generally, performing a Euclidean path integral over a manifold with boundary, with a local operator inserted somewhere in the interior, prepares a state on the boundary. This construction holds in any QFT. What fails, however, is the converse. Given a state
on a spatial sphere, one can still trace it back to some local perturbation at the origin, signalled by a divergence. However, such a singularity often cannot be matched to any local operator in the theory.\footnote{This is most easily seen by performing a coordinate transformation \(r=\ex{\tau}\), where \(\tau\) is the Euclidean time. Under this map \(\tau=-\infty\) corresponds to \(r=0\). However, in a non-conformal theory one cannot subsequently remove the conformal factor. As a result, there is an extra divergence from the metric itself which cannot, in general, be countered by a local operator.}

All hope is not lost. In CFTs what organises the state-operator correspondence is conformal symmetry. But is it possible that some other symmetry can take this organising role upon itself? In this paper, we show that this is the case. Specifically, we construct an explicit state-operator correspondence in a class of non-conformal quantum field theories that exhibit a particular global symmetry with a non-trivial 't Hooft anomaly. We will clarify the precise setting shortly.

For now, let us motivate this possibility with a familiar example: a two-dimensional (unitary) CFT with a continuous global symmetry. For simplicity take the symmetry to be \(\U(1)\) and focus on \(c=1\) CFTs. It is well known that any such a theory can be realised in terms of free fields \cite{Dotsenko:1984ad}. What's more, there is an abelian Kac--Moody algebra organising the spectrum of states and operators. In this setting, the stress tensor admits a Sugawara form \cite{Sugawara:1967rw}, meaning the generators of the conformal group (and in fact the entire Virasoro algebra) can be written in terms of the Kac--Moody modes. Hence, states and operators in this theory are naturally organised by the Kac--Moody algebra, which follows from the \(\U(1)\) global symmetry. That they also come in representations of the conformal group follows directly from the Sugawara construction.

The aim of this paper is to replicate the above situation in a more general setting, dispensing of the need for conformal invariance. The key ingredient that makes this possible is the notion of generalised global symmetries \cite{Gaiotto:2014kfa}. Over the past decade, it has become clear that the conventional idea of symmetry can be extended in multiple directions. This has led to the development of a broader framework encompassing higher-form symmetries, higher-group symmetries, and non-invertible symmetries, among others. See \cite{Sharpe:2015mja,Cordova:2022ruw,McGreevy:2022oyu,Freed:2022iao,Gomes:2023ahz,Schafer-Nameki:2023jdn,Brennan:2023mmt,Bhardwaj:2023kri,Shao:2023gho,Iqbal:2024pee,Pasternak:2025tcc} for reviews. These generalisations provide a powerful toolkit for analysing and constraining quantum field theories.

Without further ado, let us lay out our setup and briefly summarise our main results. We will consider \(d\)-dimensional quantum field theories with a zero-form symmetry and a \((d-2)\)-form form symmetry, \(\gf{\U(1)}{0}\times \gf{\U(1)}{d-2}\), that are tied together by an 't Hooft anomaly. To be precise, coupling the symmetries to background gauge fields \(\f{\cA}{1}\) and \(\f{\cB}{d-1}\), respectively we consider systems with a mixed anomaly (written here as an inflow action):
\begin{equation}\label{eq:anomaly-action}
	S_\t{anomaly} = \frac{\ii}{2\pi}\int_{M_{d+1}} \f{\cB}{d-1}\w\dd{\f{\cA}{1}}~.
\end{equation}
Oftentimes, the higher-form symmetry is not manifest in the ultraviolet but emerges in an infrared phase \cite{Delacretaz:2019brr}. Accordingly, the theories we will work with are effective field theories that realise this symmetry structure at low energies.

A first key result of this paper is that such a symmetry structure can always be realised by a superfluid effective field theory (EFT) with an appropriate choice of equation of state, which we dub \textquote{superfluidisation.} The simplest realisation is that of a single Goldstone boson: a superfluid in a minimal sense, where the equation of state dictates that the pressure scales quadratically with the chemical potential and the symmetry breaking persists even at zero charge density. This construction replaces the free field realisation of our earlier 2d CFT example and may be viewed as a higher-dimensional analogue of bosonisation. In this sense, the superfluid EFT plays the role of a universal model for capturing the anomaly and its associated charge structure. This echoes the primary result of \cite{Delacretaz:2019brr}, where it was shown that such a symmetry structure implies a massless particle in the spectrum of the theory, and resonates with the main ideas in \cite{Hinterbichler:2022agn,Hinterbichler:2024cxn} whereby the superfluid EFT was constructed from the bottom up by a particular realisation of the anomaly.

We then go on to show that the superfluid EFT possesses an \emph{infinite number of conserved charges} which we construct explicitly.\footnote{Here we use the word conserved to mean topological. Note that this is not necessarily the same as saying that they commute with the stress tensor. It is the QFT version of the statement that their \emph{total} time derivative vanishes, i.e. \(\dv{Q}{t} = \pdv{Q}{t} + \ii\,\comm{H}{Q} = 0\). As a simple example, the modes \(\alpha_n = \oint \frac{\dd{z}}{2\pi} \; z^n\, J(z)\) in the 2d free boson CFT are topological and conserved, yet they do not commute with the Hamiltonian. In other words \(z^n\) introduces explicit time dependence.} These charges are given by integrals of local currents on codimension-1 hypersurfaces:
\begin{equation}
	Q_n[\Sigma_{d-1}] = \int_{\Sigma_{d-1}} \dd[d-1]{x} J_0^{(n)}(x) = \int_{\Sigma_{d-1}} \star\f{J}{1}^{(n)}~.
\end{equation}
Here \(\f{J}{1}^{(n)}\) schematically denote modes of the \(\U(1)\) current. Similarly there are codimension-1 charges, \(\widetilde{Q}_n\), descending from the second \(\gf{\U(1)}{d-2}\) symmetry. In other words, they generate zero-form symmetries.  Upon computing their algebra, we find that these charges obey an abelian current algebra with a central extension:
\begin{equation}
	\comm{Q_n}{\widetilde{Q}_m} = \ii \sqrt{\lambda_n}\, \delta_{nm}~,
\end{equation}
where \(\sqrt{\lambda_n}\) denotes the eigenvalues of the Laplacian on \(\Sigma_{d-1}\) and \(n\) denotes the quantum numbers of the corresponding eigenfunctions. This is a direct generalisation of abelian Kac--Moody algebra to higher dimensions. In fact, it turns out that these charges are spectrum-generating. As a result the Hilbert space of states is organised into representations of this algebra. These are simply Verma modules consisting of \emph{descendants} on top of highest-weight, \emph{primary}, states. What's more, the local operators of the theory also transform under this algebra and are organised into primary and descendants.

This last fact enables our chief takeaway. Keeping the aforementioned 2d CFT example as inspiration, we adapt the methods introduced in \cite{Hofman:2024oze} to establish a state-operator correspondence in a non-conformal setup. To recall briefly, \cite{Hofman:2024oze} construed a state-operator correspondence for line operators in 4d CFTs with a continuous one-form symmetry. There it relates states on \(\S^2\times\S^1\) and line operators on \(\R^3\times\S^1\) and is governed by the current algebra \cite{Hofman:2018lfz}, not by conformal mappings, as no conformal transformation connects the relevant geometries.

Here we return to local operators, abandoning conformal invariance, and we establish the following:
\begin{center}
	\vspace{-8pt}
	In \(d\)-dimensional QFTs with a \(\gf{\U(1)}{0}\times\gf{\U(1)}{d-2}\) symmetry with anomaly \cref{eq:anomaly-action}, \\ states on \(\S^{d-1}\) are in one-to-one correspondence with local operators on \(\R^d\).
	\vspace{-8pt}
\end{center}
We construct this correspondence explicitly in two concrete realisations: Goldstone bosons and superfluid phonons. Interestingly, in both cases, we find that in the absence of conformal invariance there is a subtle difference compared to the usual state-operator correspondence for CFTs. The identity operator --- corresponding to the empty path integral --- does not prepare the vacuum state, but rather a \emph{squeezed vacuum}.

Let us emphasise that although both of our examples are free, we expect that the state-operator correspondence we construct is not tied to a free realisation. What underpins the correspondence is the current algebra, which remains intact for the full non-linear superfluid. This resonates with recent lattice approaches \cite{Berenstein:2023tru,Berenstein:2023ric,Berenstein:2024tdc} where gaplessness in the continuum, arises not from the shape of the Hamiltonian, but by the fundamental commutators --- which are a discretised version of the same abelian current algebra. Here too, the quadratic nature of the examples makes the correspondence tractable --- it allows us to construct the map between states and local operators explicitly, but it is plausibly not essential to the mechanism itself.

An organising summary is as follows. In \cref{sec:symmetries} we detail on the connection between symmetries, anomalies, and current algebras, and explain how they are realised in effective field theory. In the following couple of sections we build up towards the state-operator correspondence. We first derive it explicitly, in \cref{sec:compact-scalars}, for the case of a compact scalar in \(d\)-dimensions. Then, in \cref{sec:superfluid}, we extend our construction to the relativistic superfluid. Reconciling with \cref{sec:symmetries} this establishes the correspondence in general. We close in \cref{sec:discussion} with a discussion on applications of our formalism, as well as extensions and open questions.

\paragraph{Notation.} This paper takes place both in Lorentzian and in Euclidean signature. Whenever in Lorentzian, we use mostly plus convention for the metric. We will denote differential forms with their degree placed in as a subscript in square brackets, like \(\f{\omega}{p}\). Higher-form symmetry groups will be denoted by a superscript in square brackets, for instance: \(\gf{G}{p}\). There will be several metrics playing a role: the metric of \(d\)-dimensional spacetime, that of a fixed \((d-1)\)-dimensional Cauchy slice, and that of a \((d-1)\)-dimensional spherical slice at radius \(r\) (viewed as a radial foliation of \(d\)-dimensional flat space). The corresponding Hodge-star operators will be denoted \(\star_\s{d}\), \(\star_\s{d-1}\), and \(\star_r\), respectively.

\section{Anomalies and infinite-dimensional current algebra}\label{sec:symmetries}

Global symmetries in quantum field theory may possess an \emph{'t Hooft anomaly} \cite{tHooft:1979rat}, indicating that while the symmetry is well-defined, it cannot be consistently gauged. In a modern perspective 't Hooft anomalies are viewed as part of the symmetry data \cite{Bhardwaj:2017xup,Chang:2018iay}. Operationally, anomalies are often diagnosed by coupling the theory to a background gauge field, \(\cA\), and examining the behaviour of the partition function under background gauge transformations.\footnote{This is well suited for invertible symmetries. For non-invertible symmetries a more refined approach is needed. See e.g. \cite{Chang:2018iay,Thorngren:2019iar,Kaidi:2023maf,Zhang:2023wlu,Cordova:2023bja,Antinucci:2023ezl}.} In the presence of an anomaly, the partition function is not invariant. Rather, it picks up a phase (here, in a \(\U(1)\) example):
\begin{equation}
	\parti[\cA+\dd{\lambda}] = \exp(\ii \int_{X_d} \alpha(\cA,\lambda)) \parti[\cA]~.
\end{equation}
The phase \(\alpha(\cA,\lambda)\), modulo local counterterms, defines the anomaly and reflects the obstruction to gauging the global symmetry. A common way to encode anomalies in terms of a classical \emph{anomaly inflow} theory of the background fields in one higher dimension.

As mentioned in the introduction the systems of interest possess a zero-form \(\gf{\U(1)}{0}\) symmetry, and a \((d-2)\)-form \(\gf{\U(1)}{d-2}\) symmetry, with a mixed anomaly. In the inflow picture, the mixed anomaly is captured by the anomaly action:
\begin{equation}\label{eq:anomaly-act-main}
	S_\t{anomaly} = \frac{\ii}{2\pi}\int_{Y_{d+1}} \f{\cB}{d-1}\w\dd{\f{\cA}{1}}~.
\end{equation}
Here, \(\f{\cA}{1}\) and  \(\f{\cB}{d-1}\) are background gauge fields for the 0-form and the \((d-2)\) form symmetry, respectively. In terms of currents the anomaly manifests itself as a modification of the usual Ward identities. In particular, if the two symmetries are realised by currents \(\f{J}{1}\) and \(\f{\widetilde{J}}{d-1}\), \cref{eq:anomaly-act-main} leads to the following anomalous conservation equations:
\begin{equation}\label{eq:anomalous-ward}
	\dd\star\f{J}{1} = 0 \qq{and} \dd\star\f{\widetilde{J}}{d-1} = -\f{\cF}{2}~,
\end{equation}
where \(\f{\cF}{2}\) is the curvature of the background connection \(\f{\cA}{1}\). We follow the conventions of \cite{Hinterbichler:2024cxn} for the sign and normalisation of the anomalous Ward identity.

In most cases, the \((d-2)\)-form symmetry is not fundamental but emerges in the infrared. The simplest example realising the anomaly structure described above is a theory of Goldstone bosons in \(d\) dimensions --- a case that will be treated in detail in \cref{sec:compact-scalars}. As such, the anomalous conservation laws in \cref{eq:anomalous-ward} are generally \emph{not} constrained by anomaly matching, since the symmetry is emergent. However, as recently discussed in \cite{Seiberg:2025bqy,Gu:2025gtb}, certain ultraviolet (UV) completions may already contain part of the anomaly. In those cases a discrete subgroup of the emergent higher-form symmetry descends from an ordinary 0-form symmetry in the ultraviolet and the associated anomaly does become subject to matching constraints.

A broader class of theories realising the same symmetry structure as Goldstone bosons is provided by the effective field theory of a superfluid, as we will review. This section has a twofold purpose. First, it will be shown that in the superfluid EFT, the symmetry enhances dramatically, giving rise to an infinite tower of conserved currents. The mixed anomaly governs the algebra of these currents, resulting in a central extension that we determine. Second, this structure is shown to be generic: the anomaly can always be realised by a suitable superfluid effective theory, by a mechanism which we dub \emph{superfluidisation}.

\subsection{Infinitely many charges in the superfluid EFT}

A relativistic superfluid is a system that spontaneously breaks a \(\U(1)\) symmetry, typically interpreted as particle-number symmetry, in a finite-density state. A simple example of a UV model (in four dimensions) that leads to a superfluid phase is that of a complex scalar with quartic interaction \cite{Babichev:2018twg,Joyce:2022ydd}:
\begin{equation}\label{eq:UV-model}
	S_\t{UV} = \int \dd[4]{x} \qty(\frac{1}{2}\abs{\pd_\mu \Phi}^2 - m^2 \abs{\Phi}^2 - \lambda \abs{\Phi}^4)~.
\end{equation}
When \(m^2<0\) this is the standard regime of spontaneous symmetry breaking (SSB) giving rise to the standard Goldstone action for the phase \(\xi(x)\), of the complex scalar \(\Phi(x) = \rho(x)\, \ex{\ii \xi(x)}\). Adding a chemical potential, \(\mu\), drives the system to a finite density state. However, as was explained in \cite{Joyce:2022ydd}, even in the regime \(m^2>0\), where the vacuum of the potential sits as zero, turning on a chemical potential \(\mu>\mu_\t{crit} = m\), destabilises the vacuum, leading to SSB of the \(\U(1)\) symmetry and, again, a finite density state.

At low-energies, when the heavy fundamental degrees of freedom have been integrated out, the superfluid degrees of freedom are described by the following effective action \cite{Son:2002zn}:\footnote{The effective action holds to all orders in the coupling, though subject to higher-derivative corrections of the form \(X^m \pd^n X\), \(n\geq 1\), with \(X= \pd_\mu \xi \pd^\mu \xi\). In the regime of validity of the EFT such corrections are suppressed by powers of the UV cutoff. In certain fine-tuned EFTs where the Wilson coefficients are atypically small, these corrections may become relevant earlier that the UV scale. For a more detailed discussion, see \cite{deRham:2014wfa}.}
\begin{align}\label{eq:superfluid-action}
	S[\xi] & = \int \dd[d]{x} \p\qty\big(\partial_\mu \xi\, \partial^\mu \xi)~.
\end{align}
In the above, \(\p(\cdot)\) is a function that is smooth away from zero, related to the superfluid's equation of state. In our conventions it gives the pressure, at zero temperature, as a function of (minus) \emph{the square} of the chemical potential: \(\p = \p\qty(-\mu^2)\). The field \(\xi\) is naturally an angular variable. Canonically, we take it to be dimensionless and normalise its period to \(2\pi\). The natural scale of the problem is set by the UV cutoff scale: \(\Lambda=\abs{m}\) in the above example.

It is immediate to observe that the \(\U(1)\) symmetry is realised nonlinearly on the superfluid EFT. Shifting \(\xi\) by a constant: \(\xi\mapsto \xi+\alpha\), leaves the action invariant. Since \(\xi\) is an angular variable, so is \(\alpha\). Associated with this symmetry there is a conserved current:
\begin{equation}\label{eq:J-superfluid}
	J_\mu = \p'\qty(\partial_\mu \xi\, \partial^\mu \xi)\; \pd_\mu \xi~.
\end{equation}
The current is conserved by virtue of the equations of motion. Switching to the language of differential forms to describe conserved currents, denoting: \(\f{J}{1}= J_\mu\,\dd{x^\mu}\), its conservation equation reads:
\begin{equation}
	\dd\star\f{J}{1} = 0~.
\end{equation}
In modern terminology \cite{Gaiotto:2014kfa}, this corresponds to a conventional, or \emph{0-form}, global symmetry. As anticipated above, the superfluid EFT also exhibits a second, \emph{emergent} \((d-2)\)-form  symmetry, \(\gf{\U(1)}{d-2}\), with a \((d-1)\)-form current:
\begin{equation}
	\f{\widetilde{J}}{d-1} = \star \dd{\xi}~.
\end{equation}
This current is conserved off-shell simply by virtue of the Bianchi identities or equivalently nilpotency of the exterior derivative. As such, the corresponding symmetry is referred to as \emph{topological}. It is common to label the 0-form and $(d-2)$-form symmetries as \emph{electric} and \emph{magnetic}, respectively, reflecting the kind of matter the two currents couple to. We will use both naming conventions interchangeably throughout.

These two symmetries give rise to conserved charges in the theory. The conserved charge associated with the electric symmetry is the total \emph{particle number}:
\begin{equation}
	Q[\Sigma_{d-1}] = \int_{\Sigma_{d-1}} \star \f{J}{1}~,
\end{equation}
while the magnetic symmetry gives rise to conserved \emph{vortex number}:
\begin{equation}
	\widetilde{Q}[\gamma_1] = \int_{\gamma_1} \star \f{\widetilde{J}}{d-1}~.
\end{equation}

On top of that, a key ingredient of the superfluid EFT is that the two symmetries are tied together by an 't Hooft anomaly of the form \cref{eq:anomaly-act-main} \cite{Delacretaz:2019brr}. The anomaly is most clearly evident upon coupling the theory to a background gauge field, \(\cA_\mu\), for the zero-form \(\gf{\U(1)}{0}\) symmetry. This is done in the usual way: derivatives are traded for a gauge-covariant version thereof. Since the symmetry is realised by shifts, the covariant derivative acts on the scalar field as
\begin{equation}\label{eq:covariant-der}
	D_\mu \xi = \pd_\mu \xi - \cA_\mu~,
\end{equation}
Under this modification, the electric current, \(\f{J}{1}\) remains conserved, but \(\f{\widetilde{J}}{d-2}\) (now expressed in terms of covariant derivatives) is anomalous:
\begin{equation}\label{eq:anomaly}
	\dd\star\f{\widetilde{J}}{d-2} = - \f{\cF}{2}~.
\end{equation}

We will now take a different approach. For that we do not need the background gauge field \(\cA\), so we turn it back off. First we will show that on top of the two separately conserved currents there exists an \emph{infinite family of zero-form symmetries}. As such the associated conserved charges can have a non-trivial algebra. The anomaly is manifested as a central term in the algebra of the charges. Let us see how this comes about.

Consider a function (zero-form) \(\f{f}{0}\) and a \((d-2)\)-form \(\f{\widetilde{f}}{d-2}\), with the aid of which we define a (one-form) \emph{dressed current}, \(\cJ_f\) as:
\begin{equation}\label{eq:dressed-current}
	\star\cJ_f = \f{f}{0} \; \star \f{J}{1} + \f{\widetilde{f}}{d-2}\w\star\f{\widetilde{J}}{d-1}~.
\end{equation}
This current turns out to be conserved if the auxiliary functions are tied together by a relation. They must satisfy:
\begin{equation}\label{eq:constraint0}
	\p' \dd{\f{f}{0}} + (-1)^{d-1} \star\dd{\f{\widetilde{f}}{d-2}} = 0~.
\end{equation}
Here, it is important to stress that the appearance of \(\p'\) makes this constraint field dependent. Nonetheless, since we are examining the conservation of a current, we can use the equations of motion. Therefore, we are to evaluate \(\p'\) at a solution of the equations of motion of \cref{eq:superfluid-action}. Secondly, on non-compact spacetimes or on manifolds with a boundary, there exist infinitely many solutions to the above equation if \(\p'\) is uniformly sign-definite. Physically this corresponds to a sign-definite distribution of charge. Choosing without loss of generality positive sign, let us parametrise \(\p'=\ex{-2V}\) for some function \(V\). Simultaneously redefining for convenience \(\f{g}{0}=\ex{-V} \f{f}{0}\) and \(\f{\widetilde{g}}{2} = \ex{V} \star \f{\widetilde{f}}{d-2}\), \cref{eq:constraint0} becomes
\begin{equation}\label{eq:constraint1}
	\dd_V \f{g}{0} + \dd_V^\dagger\, \f{\widetilde{g}}{2} = 0~,
\end{equation}
where \(\dd_V = \ex{-V}\dd\, \ex{V} = \dd + \dd{V} \w \cdot\) is the Witten--Lichnerowicz differential \cite{Witten:1982im} and \(\cdd_V\) is its adjoint, \(\cdd_V = \ex{V}\cdd\ex{-V}\) with \(\cdd=(-1)^{d-1} \star\dd\star\), the codifferential (here, acting on a 2-form). So \cref{eq:constraint0} in the form of \cref{eq:constraint1} becomes a problem in the cohomology of \(\dd_V\), known as Morse--Novikov cohomology. It is a standard result \cite{Witten:1982im} that on an exact form, as is in this case \(\dd{V}\), this cohomology is isomorphic to the usual de Rham cohomology. Therefore, \cref{eq:constraint0} has the same solution space as the free \textquote{twisted self-duality} \cite{Cremmer:1997ct,Cremmer:1998px} equation:
\begin{equation}
	\dd{\f{g}{0}} + \dd^\dagger\, \f{\widetilde{g}}{2} = 0~.
\end{equation}
This case corresponds to linear pressure, or equivalently a free compact scalar, and admits infinitely many solutions that we will explicitly construct in the following section. For the time being, we rely on that general result, assume a solution and explore its consequences.

An important point to emphasise is that there is a gauge redundancy in \cref{eq:dressed-current,eq:constraint0}. In particular, the currents defined by \(\widetilde{f}\) and by \(\widetilde{f}+\dd{\f{\lambda}{d-3}}\) are equivalent; in particular they generate the same charge, as can be easily checked by integrating by parts and invoking current conservation. We work modulo such equivalences. Similarly for \(f\), we work up to constant shifts. Keeping that in mind, the counting of the degrees of freedom in \(\cJ_f\) is as follows. The function \(f\) contributes one scalar degree of freedom. The \((d-2)\)-form \(\widetilde{f}\) modulo gauge shifts, works as a \((d-2)\)-form gauge field, which carries only one polarisation --- thus contributing one additional degree of freedom. The constraint \cref{eq:constraint0} then halves the total tally. Altogether, \(\cJ_f\) is a \emph{one function worth family of currents}.

This family gives rise to a corresponding set of conserved charges supported on codimension-1 surfaces:
\begin{equation}\label{eq:charge-def}
	\cQ_f[\Sigma_{d-1}] = \int_{\Sigma_{d-1}}\star\cJ_f~.
\end{equation}
Moreover, since these charges live on codimension-1 they may have non-trivial commutators. However, since they descend from abelian symmetries, their commutators can at most be central. Indeed, the mixed anomaly \cref{eq:anomaly} between the original currents translates to a current algebra:
\begin{equation}\label{eq:current-algebra}
	\comm{J_{0}(x)}{\widetilde{J}_{i_1\cdots i_{d-1}}(y)} = \ii\, \epsilon_{i_1\cdots i_{d-1} j}\, \pd^j \delta(x-y)~.
\end{equation}
This implies the following algebra among the dressed charges:
\begin{equation}\label{eq:comm-charges}
	\comm{\cQ_f}{\cQ_h} = \ii\,(-1)^d\int_\Sigma \qty(\f{f}{0}\dd{\f{\widetilde{h}}{d-2}} - \f{h}{0}\dd{\f{\widetilde{f}}{d-2}})~.
\end{equation}
In other words we have demonstrated the existence of infinitely many conserved charges in the superfluid EFT, and we have shown that they satisfy an abelian Kac--Moody-like algebra with a central extension determined by the anomaly. In the next section we will show that this algebra is spectrum-generating and constrains the structure of the Hilbert space.

Let us close this passage with some comments. The appearance of infinitely many conserved charges points towards integrability of the superfluid EFT. In two-dimensions, where such a feature is sufficient to establish integrability, a similar result was demonstrated in \cite{Dodelson:2023uuu}. In our case, to establish integrability
we would further have to show that our dressed charges, \(\cQ_f\) imply a collection of charges, \(\cI_n\), built out of stress-tensor modes, that are in involution. This would constitute a higher-dimensional generalisation of the Kortweg--de Vries (KdV) hierarchy. While we have not done so here, we expect that it should be possible, for instance inspired by analogous two-dimensional questions \cite{Freeman:1992ag,Freeman:1993cq}. Furthermore, it is well-appreciated that Kac--Moody algebras similar to \cref{eq:comm-charges} arise in non-conformal integrable field theories (see for instance \cite{Lacroix:2023gig}). Our results tie well with both perspectives. While these observations do not constitute a proof of integrability, they offer compelling motivation for further study.

\subsection{Superfluidisation}

Let us adopt a different perspective, in the spirit of \cite{Delacretaz:2019brr}. Suppose that we do not know the precise effective field theory describing the infrared physics. Instead, we only know the global symmetries --- exact or emergent --- and their 't Hooft anomalies. Assume then, only the \(\gf{\U(1)}{0}\times\gf{\U(1)}{d-2}\) symmetry, with mixed anomaly as in \cref{eq:anomalous-ward}. We will show that such a scenario can \emph{always} be realised by an EFT of the superfluid kind, like \cref{eq:superfluid-action}. The same conclusion was reached in \cite{Hinterbichler:2024cxn} using different arguments. However, our proof connects immediately with the infinite dimensional current algebra presented above and is mirroring the mechanism of \emph{free field realisation} in two-dimensional systems.

To understand and further motivate this claim let us take a brief detour in two dimensions. In \(d=2\), the Ward identities in \cref{eq:anomalous-ward} are immediately recognisable as the \emph{axial} or \emph{chiral anomaly}, exemplified by a massless Dirac fermion coupled to an external gauge field:
\begin{equation}
	\cL = \bar{\psi}\; \ii\gamma^\mu\qty(\pd_\mu+\cA_\mu)\,\psi~.
\end{equation}
Remarkably this two-dimensional model can be described in terms of a free scalar field.\footnote{In this passage we are neglecting global features such as the dependence on spin sturctures. In a more modern approach, bosonisation refers to gauging the \((-1)^F\) symmetry of the fermionic theory. See for instance \cite{Karch:2019lnn} for a more careful exposition.} This equivalence is known as bosonisation. The main idea behind it can be traced back to the algebra satisfied by the two currents. In the fermionic theory, the currents are given by:
\begin{equation}
	J_\mu \equiv J_\mu^{\t{(vector)}} = \bar{\psi}\, \gamma_\mu \psi \qq{and} \widetilde{J}_\mu \equiv J_\mu^{\t{(axial)}} = \bar{\psi}\, \sigma_z \gamma_\mu \psi = \varepsilon_{\mu\nu} J^\nu.
\end{equation}
Owing to the simplicity of the theory one can compute directly the algebra of the two currents to find
\begin{equation}\label{eq:KM-og}
	\comm{J_0(x)}{\widetilde{J}_0(y)} = \ii \pd_x \delta(x-y)~.
\end{equation}
The bosonic interpretation becomes immediately available now. One can realise this algebra starting from a free scalar field \(\phi\). The currents are now expressed as \(J_\mu = \pd_\mu \phi\) and \(\widetilde{J}_\mu = \varepsilon_{\mu\nu}\pd^\nu{\phi}\). This is actually a much deeper statement, well understood and exploited in two-dimensional theories. A \(\U(1)\) global symmetry in any two-dimensional unitary CFT can be described in terms of a free scalar \cite{Dotsenko:1984ad}. Moreover, the symmetry enhances to a Kac--Moody algebra, governed by \cref{eq:KM-og}. There is also a non-abelian extension of the story \cite{Witten:1983ar}, realised by Wess--Zumino--Witten (WZW) models, and a higher-dimensional extension, realised by (higher-form) Maxwell CFT \cite{Hofman:2018lfz,Hofman:2024oze}.

Building on the two-dimensional example, we are in now a position to clarify and substantiate the earlier claim. We will show in the next paragraph that any theory with a \(\gf{\U(1)}{0}\times\gf{\U(1)}{d-2}\) symmetry and a mixed anomaly, as in \cref{eq:anomalous-ward}, necessarily implies that the currents satisfy the algebra:
\begin{equation}\label{eq:current-algebra2}
	\comm{J_{0}(t,\vec{x})}{\star\widetilde{J}_{i}(t,\vec{y})} = \ii\, \pd_i \delta(\vec{x}-\vec{y})~,
\end{equation}
where \(\star\widetilde{J}_{i}\) denotes the component of \(\star\widetilde{J}\) along \(\dd{x}^i\). As established in the previous section, the superfluid EFT inherently satisfies this algebra. A minimal choice is provided by taking \(\star\f{\widetilde{J}}{d-1}=\dd{\phi}\). The other current, \(\f{J}{1}\), will then necessarily have the form \cref{eq:J-superfluid} --- though this does not fix the precise form of \(\P(\cdot)\). Hence, the physics of infrared phases with this pattern of symmetries and anomalies are always realising a superfluid. From this point onward, the discussion naturally extends to the enhancement of the symmetry, encompassing infinitely many dressed charges and everything that follows from it, including our main result of the state-operator correspondence.

An elementary way to show that \cref{eq:current-algebra2} follows from \cref{eq:anomalous-ward} is to consider the path integral representation of the commutator, which reads:
\begin{equation}\label{eq:commutator}
	\begin{aligned}
		C_{i}(\vec{x},\vec{y}) & \coloneqq \ev{\cdots\comm{J_{0}(t,\vec{x})}{\star\widetilde{J}_{i}(t,\vec{y})}\cdots}                                                                                                                      \\
		                       & = \lim_{\varepsilon\to 0}\bigg(\ev{\cdots{J_{0}(x)\,\star\!\widetilde{J}_{i}(y)}\cdots}_{x^0 = y^0+\varepsilon} - \ev{\cdots{J_{0}(x)\,\star\!\widetilde{J}_{i}(y)}\cdots}_{x^0 = y^0-\varepsilon}\bigg)~,
	\end{aligned}
\end{equation}
since the path integral naturally computes time-ordered correlators. The dots denote other insertions away from \(x\) and \(y\). To compute \cref{eq:commutator} we will use the Fourier-transformed correlator \cite{Delacretaz:2019brr}, that is fixed uniquely by the mixed anomaly  :
\begin{equation}
	\Pi_{\mu\nu}(p) \coloneqq \int\dd[d]{x}\; \ex{-\ii p\cdot x} \ev{J_\nu(x)\;\star\!\widetilde{J}_\mu(0)} =  \frac{p_\mu p_\nu - \abs{p}^2 g_{\mu\nu}}{\abs{p}^2}~.
\end{equation}
The commutator is then:
\begin{equation}
	C_i(\vec{x},\vec{y}) = \lim_{\varepsilon\to 0}\int\frac{\dd[d]{p}}{(2\pi)^{d}}\ \Pi_{0i}(p) \qty(\ex{-\ii p_0 \varepsilon + \ii\vec{p}\cdot\qty(\vec{x}-\vec{y})} - \ex{\ii p_0 \varepsilon + \ii\vec{p}\cdot\qty(\vec{x}-\vec{y})})~.
\end{equation}
Upon performing the integral over \(p_0\) (regulating the pole at \(p_0 = \abs{p}\) by an \(\ii \epsilon\) prescription) the remaining integrand is smooth and we can take the limit \(\varepsilon\to 0\), finding finally:
\begin{equation}
	C_i(\vec{x},\vec{y}) = \int \frac{\dd[d-1]{\vec{p}}}{(2\pi)^{d-1}} \ex{\ii \vec{p}\cdot(\vec{x}-\vec{y})}\ \ii\,p_i~.
\end{equation}
The last equation we immediately recognise as the Fourier transform of \(\partial_i \delta(\vec{x}-\vec{y})\), hence establishing \cref{eq:current-algebra2}.

\section{Compact scalars in \texorpdfstring{\(d\)}{d} dimensions}\label{sec:compact-scalars}

In this section we construct the advertised state-operator correspondence in the simplest possible scenario: a free compact scalar in \(d\) dimensions. It is connected to the superfluid by choosing a linear function for the pressure. In this case the EFT is analytic at zero chemical potential and can be arrived at, for instance, from the \(m^2<0\) regime of the above UV model \cref{eq:UV-model}. The action reads:
\begin{equation}
	S[\phi] = \frac{\g}{2}\int_X \dd[d]{x} \pd_\mu \phi\, \pd^\mu \phi = \frac{\g}{2}\int_X\dd{\phi}\w\star\dd{\phi}~.
\end{equation}
Our conventions are such that \(\phi\) is dimensionless and its periodicity is \(2\pi\). Consequently the coupling \(\g\) has dimension \(d-2\). It sets the scale of validity of the Goldstone EFT (sometimes we will use the actual energy scale \(\Lambda = \g^{\frac{1}{d-2}}\)).

It is important to stress that the theory, despite free, is not conformal in \(d>2\).\footnote{In the present paper we will focus on \(d>2\) throughout. In \(d=2\) the theory is conformal and the state-operator correspondence is textbook material. For the more subtle case of \(d=1\), i.e. quantum mechanics, see \cite{Sen:2011cn}.} The sharpest indication comes from the existence of local vertex operators, \(V_p(x) = {\ex{\ii p \phi(x)}}\), with \(p\in\Z\) whose two-point function behaves as\footnote{Since the theory is free this follows immediately by the two-point function of \(\phi\), \(\ev{\phi(x)\phi(y)} = \frac{1}{\g\abs{x-y}^{d-2}}\) (up to gauge-dependent terms having to do with the fact that \(\phi\) is compact).}
\begin{equation}\label{eq:2pt-fn-Vp}
	\ev{V_p(x)V_{-p}(y)} = \exp(\frac{p^2}{\g\abs{x-y}^{d-2}})~.
\end{equation}
This is incompatible with conformal field theory, unless \(d=2\) where it becomes a standard power law. There are two conformal fixed points of this theory. One sits at \(\g\to 0\), corresponding to a UV fixed point. It is a free, scale invariant, but not conformal theory \cite{El-Showk:2011xbs,Nakayama:2013is,Dymarsky:2013pqa,Lee:2021obi}: a free scalar with its zero mode removed. In the infrared (\(\g\to\infty\)) the scalar decompactifies and it becomes a bona fide CFT.\footnote{On a spacetime with non-vanishing curvature one needs to turn on the conformal coupling \(\sim \cR \phi^2\) to preserve conformal invariance. Note that this couplng is only allowed at \(\g\to\infty\), since at any other value it is incompatible with the scalar's periodicity.} In the current work we will study the theory at an intermediate point, at finite coupling.

Let us begin analysing the symmetries. The story follows immediately from the general discussion in \cref{sec:symmetries}. The model enjoys a \(\gf{\U(1)}{0}\times\gf{\U(1)}{d-2}\) symmetry with currents
\begin{equation}
	\f{J}{1} = \dd{\phi}\qquad \qq{and}\qquad \f{\widetilde{J}}{d-1} = \star\dd{\phi}~,
\end{equation}
and anomaly \cref{eq:anomaly-act-main}. Identically as in \cref{sec:symmetries}, this implies the conservation of an infinite family of dressed currents \(\cJ_f\) defined as in \cref{eq:dressed-current}. This time the dressing functions are linked by a \textquote{twisted self-duality} \cite{Cremmer:1997ct,Cremmer:1998px} constraint
\begin{equation}\label{eq:constraint-free}
	\dd{\f{f}{0}} + (-1)^{d-1} \star\dd{\f{\widetilde{f}}{d-2}} = 0~.
\end{equation}
Continuing on with this line of reasoning, this gives infinitely many conserved charges, \(\cQ_f[\Sigma_{d-1}]\), as in \cref{eq:charge-def} satisfying again \cref{eq:comm-charges}.

As stated above, \cref{eq:comm-charges} is equivalent to a \(\widehat{\fu}(1)\) Kac--Moody algebra. This equivalence becomes more transparent when we expand the relevant currents and functions in terms of a suitable basis. A natural choice is the basis formed by the the eigenfunctions and eigen-\((d-2)\)-forms of the Laplacian on \(\Sigma\). For the \((d-2)\)-forms we find it convenient to resolve the gauge redundancy described above by working in Coulomb gauge \(\csd\widetilde{f}=0\), where the bold differential denotes the exterior derivative on \(\Sigma\). By Poincaré duality, except for the zero-modes, the eigen-\((d-2)\)-forms are completely specified by the scalar eigenfunctions.

To see why, let \(y_n\) be an eigenfunction of the Laplacian with eigenvalue \(\lambda_n\neq 0\). Then the \((d-2)\)-form
\begin{equation}\label{eq:beltramisation}
	\widetilde{y}_n = \frac{(-1)^d}{\sqrt{\lambda_n}} \star_\s{d-1} \sd{y_n}
\end{equation}
is an eigenform of the transversal Laplacian. Here, \(\star_\s{d-1}\) denotes the Hodge star on \(\Sigma_{d-1}\). All non-zero modes can be generated this way.\footnote{In \(d=2\) \cref{eq:beltramisation} leads to the holomorphic/anti-holomotphic split of the compact scalar CFT.} The factor of \(\lambda_n^{-1/2}\) ensures that \(\widetilde{y}_n\) is properly normalised, assuming \(y_n\) itself is normalised, while the dimension-dependent sign is simply for future convenience.

Together with the scalar zero-mode \(y_0=1/\sqrt{\vol(\Sigma)}\) and the harmonic \((d-2)\)-forms \(\widetilde{y}_{0i}\), \(i=1,\cdots b_1(\Sigma)\) (with \(b_1\) the first Betti number) this construction provides a complete basis. Throughout this discussion we are assuming \(\Sigma\) to be compact and connected, although these assumptions can be relaxed without difficulty.

Altogether the currents are expanded as:
\begin{align}\label{eq:current-mode-exp}
	\begin{aligned}
		\eval{\qty(\star_{\scriptscriptstyle d} \f{J}{1})}_{\Sigma}               & = Q_{0}  \star_\s{d-1} y_0 + \sum_{n} Q_n \star_\s{d-1} y_n~,                                                                           \\
		\eval{\qty(\star_{\scriptscriptstyle d} \f{\widetilde{J}}{d-1})}_{\Sigma} & = \sum_{i=1}^{b_1(\Sigma)} \widetilde{Q}_{0j} \star_\s{d-1}\widetilde{y}_{0j} + \sum_{n} \widetilde{Q}_n \star_\s{d-1}\widetilde{y}_n~,
	\end{aligned}
\end{align}
The zero modes, \(Q_0\), \(\widetilde{Q}_{0j}\) are the original \(\gf{\U(1)}{0}\times\gf{\U(1)}{d-2}\) momentum and winding charges. Their commutators are of course trivial. The rest of the modes, \(Q_n\) and \(\widetilde{Q}_n\), are the novel conserved charges following from \cref{eq:dressed-current,eq:charge-def}. Their algebra follows from \cref{eq:comm-charges} and reads:
\begin{equation}\label{eq:KM-modes}
	\comm{Q_n}{\widetilde{Q}_m} = \frac{\ii}{\g}\, \sqrt{\lambda_n} \delta_{nm}~.
\end{equation}
We stress here that \(n\) is not an integer but a collection of indices: the quantum numbers of the Laplacian on \(\Sigma\). For instance, if \(\Sigma\) is a \((d-1)\)-torus, \(n\) is a \((d-1)\)-tuple of integers \(n=(k_1,\cdots,k_{d-1})\) labelling the momenta along each cycle of the torus, while if \(\Sigma\) is a sphere \(n=(\ell,m_1,\cdots,m_{d-2})\) is the total angular momentum and its various projections. Hence the precise form of the algebra depends on \(\Sigma\). In contrast, in \(d=2\) there is only one reasonable choice of spatial slice --- a circle --- making the 2d Kac--Moody algebra essentially unique.

Equation \cref{eq:KM-modes} describes the mode algebra of the currents, in a way that naturally generalises Kac--Moody algebras commonly appearing in two-dimensional CFTs. We note here a few variations of this algebra that have recently appeared in the literature. In \cite{Hofman:2018lfz} a higher-form version of \cref{eq:comm-charges} was identified in higher-dimensional CFTs. Expanded in modes \cite{Fliss:2023uiv,Hofman:2024oze}, the algebra of \cite{Hofman:2018lfz} is structurally identical to \cref{eq:KM-modes}, although the physical interpretation is slightly different. In \cite{Fliss:2023uiv} this algebra organises gapless \emph{edge modes} of topological field theories, while in \cite{Hofman:2024oze} the setup was higher-dimensional conformal field theory.

\subsection{The states}\label{ssec:scalar-states}

We now have identified all the relevant symmetries for our story. The next step towards establishing a state-operator correspondence is to construct the space of states. While this is straightforward given the quadratic nature of the theory, it is helpful to do so in a way that makes the connection to the symmetry considerations of the previous section explicit. To that end, we now consider \(\Sigma\) as the Cauchy slice on which the states live. Specifically, we take \(\Sigma\) to be a \((d-1)\)-dimensional sphere of radius \(R\): \(\Sigma_R \equiv \S^{d-1}_R\). This choice turns out to be the appropriate Cauchy slice for defining the Hilbert space of local operators. In the non-conformal case, this is not immediately obvious, as there is no Weyl equivalence argument that uniquely selects this slice --- nonetheless it proves to be the correct one. We will comment on other choices of slices in the discussion.

The Hamiltonian can be written in a Sugawara form in terms of the two currents:
\begin{equation}\label{eq:Hamiltonian-free}
	H = \frac{\g}{2} \qty(\norm{J_{\Sigma_R}}^2+\norm*{\widetilde{J}_{\Sigma_R}}{\!\vphantom{\norm{J_\Sigma}}}^2)~,
\end{equation}
where the subscript denotes restriction on \(\Sigma_R\) and the norm is with respect to the standard Hodge inner product \(\int_\Sigma \;\cdot\;\w\star\ \cdot\;\). This is simply the standard form \(\sim \Pi^2 + \qty(\nabla \phi)^2\). The benefit of using this language to express the Hamiltonian is that mode expanding the currents in spherical harmonics, like in \cref{eq:current-mode-exp}, immediately reveals a countable collection of harmonic oscillators:
\begin{equation}
	H_\Sigma = \frac{\g}{2}\; Q_0^2
	+ \sum_{\ell,\vec{m}} A_\vlm^\dagger A^{\phantom{\dagger}}_\vlm~,
\end{equation}
labelled by angular momentum, \(\ell\in\Z_{>0}\), and a vector of integers \(\vec{m} = (m_1,\cdots,m_{d-2})\) satisfying
\begin{equation}\label{eq:l-range}
	\ell \geq m_1 \geq \cdots m_{d-3}\geq \abs{m_{d-2}}~.
\end{equation}
These are really the quantum numbers of spherical harmonics on \(\S^{d-1}\). See \cref{app:spherical harmonics} for details. Here we have excluded the mode \(\ell=0\) as it is contained in the zero-mode part. In the above we have identified ladder operators:
\begin{align}\label{eq:ladder-def}
	A_\vlm & = \sqrt{\frac{\g}{2}}\qty(Q_\vlm+\ii\,\widetilde{Q}_\vlm)~,
\end{align}
with daggers obtained by hermitian conjugation, and we have also neglected a (possibly infinite) normal ordering constant. The modes \(A_\vlm^{(\dagger)}\) satisfy the usual algebra of creation and annihilation operators:
\begin{equation}\label{eq:ladder-algebra}
	\comm{A_\vlm^{\phantom{\dagger}}}{A^\dagger_{\ell'\vec{m}'}} = \sqrt{\lambda_{\ell}\,}\  \delta_{\ell,\ell'}\delta_{{\vec{m},\vec{m}'}}~.
\end{equation}
From the Hamiltonian one sees immediately that \(A_\vlm^\dagger\) raises the energy by
\begin{equation}
	\sqrt{\lambda_{\ell}\,} = \frac{\sqrt{\ell(\ell+d-2)}}{R}
\end{equation}
units and \(A_\vlm\) lowers it by the same amount. Finally the zero mode, \(Q_{0}\), commutes with everything. Note that since \(\Sigma \cong \S^{d-1}\) has trivial first homology in \(d>2\) there are no states charged under the \((d-2)\)-form symmetry.

The Hilbert space on \(\Sigma_R\) splits into superselection sectors labelled by the electric charge, which is quantised according to
\begin{equation}\label{eq:0form-flux-quant}
	\int_{\Sigma_R}\star_{\scriptscriptstyle d}\f{J}{1} = \frac{p}{\g}~, \qquad p\in\Z~.
\end{equation}
This follows essentially from dualising the theory to a \((d-2)\)-form gauge field and inferring magnetic flux quantisation. See also \cite{Witten:1979ey,Verlinde:1995mz}. In other words the Hilbert space is graded as
\begin{equation}
	\cH_{\Sigma_R} = \bigoplus_{p\in\Z} \cH_p~.
\end{equation}

The first states to populate each sector are \emph{primary} or \emph{highest-weight states} of the underlying Kac--Moody algebra. Explicitly, these are states of fixed charge \(\ket{p}\) that are annihilated by all lowering operators:
\begin{equation}
	\begin{aligned}\label{eq:primary-conditions}
		Q_0 \ket{p}    & = \frac{p}{\g\,\sqrt{\vol(\Sigma)}}\ket{p} = \frac{p}{\g\, \V^{1/2}\; R^{\frac{d-1}{2}}}\ket{p}~, \\[0.7em]
		A_\vlm \ket{p} & = 0 \qquad \t{for all \(\ell\) and \(\vec{m}\)}~.
	\end{aligned}
\end{equation}
In the above, \(\V\) is the volume of a unit \((d-1)\)-sphere. The energy of the primary states is
\begin{equation}\label{eq:primary-energy}
	\Delta_p = \frac{p^2}{2\g\,\V\, R^{d-1}}~.
\end{equation}

On top of these states, each sector is populated by \emph{descendant} or excited states by acting with creation operators. In total each superselection sector is spanned by states of the form
\begin{equation}\label{eq:states-sphere}
	\ket{p;\set{N_\vlm}} \coloneqq \prod_{\ell,\vec{m}} \qty(A_\vlm^\dagger)^{N_\vlm}\ket{p}~,
\end{equation}
with \(N_\vlm\in\Z_{\geq 0}\), with energy
\begin{equation}
	E_{\set{N_\vlm}} = \Delta_p+\frac{1}{R}\sum_{\ell,\vec{m}} N_\vlm \sqrt{\ell(\ell+d-2)}~.
\end{equation}
It was emphasised in \cite{Fliss:2023uiv,Hofman:2024oze} that this is precisely the structure of Verma modules of the extended current algebra we identified above. It is essentially this structure that enables the construction of the state-operator correspondence as we will show immediately.

Let us make a few clarifying remarks. The underlying theory is an effective theory of Goldstone bosons, valid when the IR scale \(1/R\) set by the radius of the sphere is well below the UV cutoff \(\Lambda\):
\begin{equation}\label{eq:scales}
	\frac{1}{R} \ll \Lambda~.
\end{equation}
In \(d>2\), this implies a parametrically small gap
\begin{equation}
	\Delta \sim \frac{1}{\Lambda^{d-2} R^{d-1}} \ll \frac{1}{R}~.
\end{equation}
so that on a large but finite spatial sphere the primary states sit isolated from the first excited states. By contrast, in \(d=2\) one finds \(\Delta\sim 1/R\), precluding spontaneous breaking consistent with the Coleman--Mermin--Wagner theorem \cite{Mermin:1966fe,Coleman:1973ci}. In the strict infinite-volume limit \(\Lambda R\to\infty\), symmetry breaking occurs and genuine Goldstone modes emerge. A convenient way to isolate them is to introduce a small temperature \cite{Maeda:2025ycr}, \(1/\beta\), such that
\begin{equation}
	\frac{1}{\Lambda^{d-2} R^{d-1}} \ll \frac{1}{\beta}\ll \frac{1}{R}~.
\end{equation}
This has the effect of damping the excited states, as they are accounted for by \(\ex{-\beta E_{\set{N_\vlm}}}\) in the canonical partition function. The remaining states, the primaries \(\ket{p}\), become effectively degenerate vacua, which can be reassembled in the usual symmetry breaking vacua:
\begin{equation}
	\ket{\theta} = \sum_{p\in\Z} \ex{\ii p \theta} \ket{p}~, \qquad
	U_\alpha \ket{\theta}=\ket{\theta+\alpha}~.
\end{equation}

At finite volume symmetry restoration takes place. All the vacua except for one get lifted, leaving a unique ground state, \(\ket{p=0}\), and a discrete spectrum of excitations. The rest of the primary states have finite energy and a parametrically large gap. Do notice, however, that for EFT to make sense at finite volume, one should not excite too many modes; one should formally truncate the space of states at some \(\abs{p}\leq p_\t{max}\) and \(\ell\leq \ell_\t{max}\), such that the energy of the corresponding state is well below the cutoff. In the following we will keep all the states to show how they are matched by the local operators. This could either be interpreted in an infinite volume limit, or that one trusts only a finite number of states/operators before UV physics kicks in.

\subsection{The operators}\label{ssec:scalar-operators}

The next ingredient to construct the state-operator correspondence lies in identifying the set of operators at play. The details are in fact very similar to how one constructs explicitly the correspondence in the 2d case, see for instance \cite{Tong:2009np,Hofman:2024oze}.

The compact scalar itself is not a well-defined local operator as it is not gauge-invariant under \(\phi\sim\phi+2\pi\). The remedy is standard; we have to consider exponentials thereof. These are known as \emph{vertex operators}:
\begin{equation}
	V_p(x) = {\ex{\ii p \phi(x)}}~.
\end{equation}
In the above, \(p\) is an integer as dictated by the periodicity of \(\phi\). The vertex operators will play an important role in what follows. The main reason is that they have fixed charge under the zero-form \(\gf{U(1)}{0}\) symmetry. Explicitly, if \(\sigma\) is a codimension-1 surface surrounding \(x\), the action of
\begin{equation}
	U_\alpha\qty[\sigma] \coloneqq \exp(\ii\, \g\,\alpha \int_\sigma \star \f{J}{1})~,
\end{equation}
on \(V_p(x)\) is
\begin{equation}\label{eq:U-Vp}
	U_\alpha\qty[\sigma] \cdot V_p(x) = \ex{\ii\alpha p}\; V_p(x)~.
\end{equation}

Another way to construct gauge-invariant operators out of the fundamental field, \(\phi\), is by taking derivatives. In other words, another class of well-defined local operators is given by the currents
\begin{equation}\label{eq:current-ops}
	J(x) \qq{and} \widetilde{J}(x)~,
\end{equation}
and derivatives thereof
\begin{equation}\label{eq:derivative-ops}
	\zeta^{\mu\nu} \partial_\mu J_\nu(x)~, \qquad \zeta^{\mu\nu_1\cdots\nu_{d-2}}\partial_\mu \widetilde{J}_{\nu_1\cdots\nu_{d-2}}(x)~, \qq{etc.}
\end{equation}
Together with the vertex operators these provide a basis of local operators. An arbitrary local operator can be constructed out of normal-ordered products of currents, their derivatives and the vertex operators. This is, for instance, the standard way one obtains the operators corresponding to physical excitations: photons, gravitons, etc. in string theory.

An equivalent way to express all local operators, that will turn out to be more useful for us is the following. Consider a small (codimension-1) surface \(S(\varepsilon,x)\) of size \(\varepsilon\) surrounding a point \(x\). Integrating \(J\) and \(\widetilde{J}\) against arbitrary smooth functions and taking the limit \(\varepsilon\to 0\) gives an infinite collection of gauge-invariant local operators
\begin{equation}
	\begin{aligned}\label{eq:descendants}
		\cD_\alpha(x) & = \lim_{\varepsilon\to 0} \int_{S(\varepsilon,x)} \f{\alpha}{d-2} \w \f{J}{1}~, \qq{and} \\ \widetilde{\cD}_{\beta}(x) &= \lim_{\varepsilon\to 0} \int_{S(\varepsilon,x)} \f{\beta}{0} \w \f{\widetilde{J}}{d-1}~.
	\end{aligned}
\end{equation}
Of course these operators contain exactly the same information as \cref{eq:current-ops,eq:derivative-ops} as can be seen by Taylor expanding the functions \(\alpha\) and \(\beta\). The benefit of working in this approach is that we can expand the dressing functions in a complete basis of functions on \(S(\varepsilon,x)\), as we have done in \cref{ssec:scalar-states} and end up with a countable basis of our local operators. Taking \(S(\varepsilon,x)\) to be a \((d-1)\)-sphere is sufficient. The resulting operators are then also labelled by the angular momenta \(\ell\) and \(\vec{m}\):
\begin{equation}
	\cD_\vlm(x) \qq{and} \widetilde{\cD}_\vlm(x)~.
\end{equation}
Again, normal-ordered products of \(\cD_\vlm(x)\), \(\widetilde{\cD}_\vlm(x)\) and \(V_p(x)\) generate all local operators of the theory.

We mention just for completeness that besides local operators, the theory also contains \((d-2)\)-dimensional disorder operators. They are obtained by removing a \((d-2)\)-dimensional locus from spacetime and prescribing boundary conditions fixing the flux around a line \(\gamma\) linking with it:
\begin{equation}
	\int_{\gamma_1} J = 2\pi w~, \qquad w\in\Z~.
\end{equation}
These operators will not play a role in the present discussion. We therefore defer their mention to \cref{sec:discussion} where we will discuss extensions of our results to  nonlocal operators.

\subsection{The correspondence}
\renewcommand{\r}{r}
To arrive at the correspondence we have in mind the following setup. Consider placing a local operator \(\cO\) somewhere in spacetime and perform a Euclidean path integral on a ball of radius \(R\) centred around that point. This will prepare a state
\begin{equation}\label{eq:state-path-int}
	\ket{\cO} = \int_{\B^d_R} \DD{\phi} \ex{-S[\phi]}\, \cO(0)~,
\end{equation}
in the Hilbert space over \(\pd\B^d_R = \S^{d-1}_R\). In writing \cref{eq:state-path-int} we are abusing notation. One should impose boundary conditions, say \(\phi=\phi_\s{\pd}\) on \(\pd\B^d_R\). The path integral with these boundary conditions will then prepare a wavefunctional of \(\phi_\s{\pd}\)
\begin{equation}
	\Psi_\cO\qty[\phi_\s{\pd}] = \braket{\phi_\s{\pd}}{\cO}~.
\end{equation}
We will nevertheless stick with \cref{eq:state-path-int}, keeping in mind the precise interpretation.

The constraints imposed by the Kac--Moody algebra \cref{eq:KM-modes} are stringent enough to allow us to compare the states \(\ket{\cO}\) to the states we obtained previously, by canonically quantising the theory. In particular we will now show that \emph{all} states can be obtained this way. However, as we will demonstrate below, the price to pay for the lack of Weyl invariance is that the states created by \textquote{simple} operators --- those involving only finitely many derivatives or modes --- will, in general, correspond to \emph{squeezed states} with respect to the Kac--Moody algebra \cref{eq:KM-modes}.

To build intuition, let us first present an informal version of the argument before refining it in the following paragraphs. The vertex operators \(V_p(x)\) carry fixed global \(\U(1)\) charge. Therefore we expect that the states they prepare,
\begin{equation}\label{eq:Vp-states}
	\ket{V_p} = \int_{\B^d_R} \DD{\phi} \ex{-S[\phi]}\, V_p(0)~,
\end{equation}
will be states of fixed charge. These states belong in the superselection sector \(\cH_p\).  Moreover, since the symmetry is abelian, the operators \(\cD_\vlm(x)\) and \(\widetilde{\cD}_\vlm(x)\) in \cref{eq:descendants} constructed entirely out of the two currents do not alter the \(\U(1)\) charge of the corresponding state. In other words,
\begin{equation}\label{eq:states-rough}
	\ket{V_p},\ket{\cD_\vlm V_p}, \ket{\widetilde{\cD}_\vlm V_p}, \ket{\cD_{\ell\vm}\widetilde{\cD}_{\ell'\vm'} \cdots V_p}\cdots \in\cH_p~.
\end{equation}
However, not all such states are expected to be linearly independent. A simple counting reveals twice as many modes as there are states in the Hilbert space.\footnote{To make the counting precise, since the Hilbert space is infinite-dimensional, one has to truncate the mode expansion up to some arbitrary \(\ell = \ell_\t{max}\) and similarly the states up to level \(\ell_\t{max}\). For any such cutoff, there are always twice as many modes as there are independent states.} Nonetheless the ladder operators we introduced in \cref{ssec:scalar-states} relate states across different levels. This ensures that states with a different number of insertions will be linearly independent. Therefore the states \cref{eq:states-rough} suffice to populate the entire superselection sector \(\cH_p\). Repeating the same argument across all superselection sectors reconstructs the entire Hilbert space, thereby establishing the desired state-operator correspondence.

That said, a few subtleties remain. First, we are still to identify exactly which are the linearly independent states on each level. Second, there is the question of energies. On a finite ball, there is no time-translation symmetry, and consequently, no conserved Hamiltonian. Instead, the analogue of time evolution --- defined via the Euclidean path integral \cref{eq:state-path-int} --- is \emph{radial evolution}. In a CFT this is generated by the dilatation operator, which is conserved. As a result, the states prepared by the path integral have fixed scaling weight, which is related to their energy by the cylinder Hamiltonian by a constant zero-point shift induced by the Weyl transformation. In contrast, for a non-conformal theory as is our case, the dilatation operator is not conserved, and the states prepared by radial quantisation do not have fixed energies. Put differently, annihilation operators on the cylinder become an admixture of creation and annihilation operators in the interior of the ball. Luckily, the situation is not troublesome as there is a standard way to deal with it: one simply needs dress the operators by a \emph{squeezing operator} that undoes the Bogoliubov transformation. In what follows, we make these arguments precise and resolve all remaining subtleties.

\begin{figure}[tb]
	\def\svgwidth{\linewidth}
\begingroup%
  \makeatletter%
  \providecommand\color[2][]{%
    \errmessage{(Inkscape) Color is used for the text in Inkscape, but the package 'color.sty' is not loaded}%
    \renewcommand\color[2][]{}%
  }%
  \providecommand\transparent[1]{%
    \errmessage{(Inkscape) Transparency is used (non-zero) for the text in Inkscape, but the package 'transparent.sty' is not loaded}%
    \renewcommand\transparent[1]{}%
  }%
  \providecommand\rotatebox[2]{#2}%
  \newcommand*\fsize{\dimexpr\f@size pt\relax}%
  \newcommand*\lineheight[1]{\fontsize{\fsize}{#1\fsize}\selectfont}%
  \ifx\svgwidth\undefined%
    \setlength{\unitlength}{1110.33082977bp}%
    \ifx\svgscale\undefined%
      \relax%
    \else%
      \setlength{\unitlength}{\unitlength * \real{\svgscale}}%
    \fi%
  \else%
    \setlength{\unitlength}{\svgwidth}%
  \fi%
  \global\let\svgwidth\undefined%
  \global\let\svgscale\undefined%
  \makeatother%
  \begin{picture}(1,0.2249669)%
    \lineheight{1}%
    \setlength\tabcolsep{0pt}%
    \put(0,0){\includegraphics[width=\unitlength,page=1]{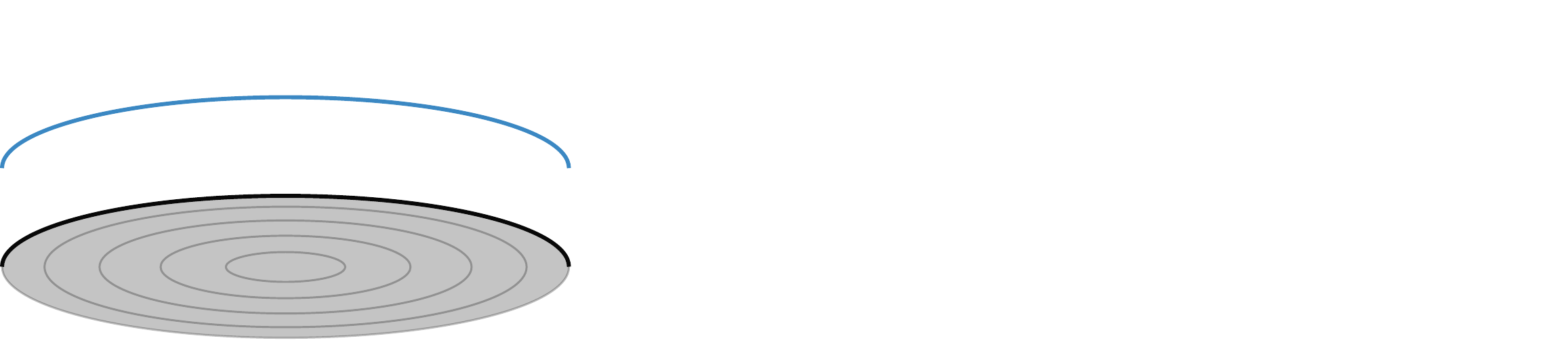}}%
    \put(0.4774179,0.10198248){\color[rgb]{0,0,0}\makebox(0,0)[lt]{\lineheight{1.25}\smash{\begin{tabular}[t]{l}\scalebox{1.4}{$=$}\end{tabular}}}}%
    \put(0,0){\includegraphics[width=\unitlength,page=2]{Qaction.pdf}}%
    \put(0.20018042,0.06429884){\color[rgb]{0,0,0}\makebox(0,0)[lt]{\lineheight{1.25}\smash{\begin{tabular}[t]{l}$\cO(0)$\end{tabular}}}}%
    \put(0,0){\includegraphics[width=\unitlength,page=3]{Qaction.pdf}}%
    \put(0.0271273,0.05637686){\color[rgb]{0.23137255,0.53333333,0.76470588}\makebox(0,0)[lt]{\lineheight{1.25}\smash{\begin{tabular}[t]{l}$Q$\end{tabular}}}}%
    \put(0,0){\includegraphics[width=\unitlength,page=4]{Qaction.pdf}}%
    \put(0.75487681,0.00391834){\color[rgb]{0.23137255,0.53333333,0.76470588}\makebox(0,0)[lt]{\lineheight{1.25}\smash{\begin{tabular}[t]{l}$Q$\end{tabular}}}}%
    \put(0.83593233,0.06429884){\color[rgb]{0,0,0}\makebox(0,0)[lt]{\lineheight{1.25}\smash{\begin{tabular}[t]{l}$\cO(0)$\end{tabular}}}}%
  \end{picture}%
\endgroup%

	\caption{Charges act on a state created by a local operator, \(\cO\),  by radially evolving inwards and linking with \(\cO\).}
	\label{fig:charges-act}
\end{figure}

\subsubsection*{Radial evolution done right}

To see how charges act on states prepared by local operators, it is necessary to track their radial evolution inward, towards the centre of the ball, as illustrated in \cref{fig:charges-act}. This is achieved by solving the constraint equation \cref{eq:constraint-free}, which ensures conservation of the dressed charges. In practice, this amounts to solving the conservation equations for the original currents \(\f{J}{1}\) and \(\f{\widetilde{J}}{d-1}\) inside the ball. This, in turn, yields explicit expressions to
\begin{equation}
	\cQ_f = \int_{\Sigma_r} \qty(\f{f}{0}(r) \; \star \f{J}{1} + \f{\widetilde{f}}{d-2}(r)\w\star\f{\widetilde{J}}{d-1})~.
\end{equation}
where the functions \(\f{f}{0}(r)\) and \(\f{\widetilde{f}}{d-2}(r)\) are now fully determined. \(\Sigma_r=\S^{d-1}_r\) denotes a \((d-1)\)-sphere at radius \(r\in\closed{0}{R}\).\footnote{Here it is understood that we consider the ball as foliated by spheres. Explicitly the metric of the ball is \(\dd{s}^2 = \dd{\r^2} + \r^2\dd{\Omega^2_{d-1}}\) with \(r\in\closed{0}{R}\).}

This is essentially all we need to determine the states. To illustrate, consider a state, \(\ket{\cO}\), prepared by the local operator \(\cO\). Acting on it with the dressed charge \(\cQ_f\) gives
\begin{equation}\label{eq:Qf-action}
	\cQ_f \ket{\cO} = \lim_{r\to 0} \int_{\B^d_R} \DD{\phi} \ex{-S[\phi]} \int_{\Sigma_r} \qty(\f{f}{0}(r) \; \star \f{J}{1} + \f{\widetilde{f}}{d-2}(r)\w\star\f{\widetilde{J}}{d-1})\times \cO(0)~.
\end{equation}
In this form we can perform the operator product expansion (OPE) of the currents with \(\cO\). Importantly, this does \emph{not} require conformal invariance as we are taking the strict \(r\to 0\) limit \cite{Hollands:2023txn}. After the OPE, the remaining integral over \(\Sigma_r\) can be evaluated explicitly, yielding the new state, \(\ket{\cO'}\). This is what is really happening under the hood in \cref{fig:charges-act}.

The technical part of the argument begins by solving the radial evolution. As mentioned above, this amounts to solving the current conservation equations, alongside with the duality equation \(\widetilde{J} = \ii \star_\s{d} J\).\footnote{The extra factor of \(\ii\) arises from working in Euclidean signature.}  It is useful to decompose the two currents in a radial part and a tangential part as
\begin{equation}
	\f{J}{1} = \dd{\r}\;J_\r + J_{\Sigma_\r} \qq{and} \f{\widetilde{J}}{d-1} = \dd{\r}\w \widetilde{J}_\r + \widetilde{J}_{\Sigma_\r}
\end{equation}
where \(J_\r\in \Omega^0\qty(\Sigma_\r)\), \(J_{\Sigma_\r}\in \Omega^1\qty(\Sigma_\r)\), \(\widetilde{J}_\r\in \Omega^{d-2}\qty(\Sigma_\r)\), and \(\widetilde{J}_{\Sigma_\r}\in \Omega^{d-1}\qty(\Sigma_\r)\). In this decomposition the conservation of the two currents implies a radial evolution equation:
\begin{equation}\label{eq:RE}
	\begin{aligned}
		\pd_\r J_{\Sigma_\r} + \ii\, \sd\star_\r\widetilde{J}_{\Sigma_\r}  & = 0~, \\
		\pd_\r \widetilde{J}_{\Sigma_\r} + \ii\, \sd\star_\r J_{\Sigma_\r} & = 0~,
	\end{aligned}
\end{equation}
together with Gauss laws:
\begin{equation}
	\sd J_{\Sigma_\r} = 0 \qq{and} \sd \widetilde{J}_{\Sigma_\r} = 0~.
\end{equation}
In the above, \(\star_r\) denotes the Hodge-star on \(\Sigma_r\), stressing the dependence on the radius. Note that the radial components of the two currents were eliminated by the duality equation. The boundary condition for the radial evolution is such that the Euclidean ball is smoothly glued on the Lorentzian cylinder, \(\R\times\S^{d-1}_R\). This is equivalent to demanding that at \(r=R\) the two currents take the form \cref{eq:current-mode-exp}.

The above boundary value problem can be readily solved (details provided in \cref{app:radial-evol}) to yield:
\begin{equation}\label{eq:currents-solved}
	\begin{aligned}
		\eval{\qty(\star_{\scriptscriptstyle d} \f{J}{1})}_{\Sigma_r}               & = \frac{Q_0 R^{\frac{d-1}{2}}}{\V^{1/2}}\;\dd{\Omega_{d-1}}                                                                                                                                              \\
		                                                                            & +\frac{1}{\sqrt{2}} \sum_{\ell,\vec{m}} \sech(2 v_\ell)\qty(\ex{-v_\ell} C_\vlm\, \qty(\frac{r}{R})^{\delta_\ell^+}+\ex{v_\ell} B_\vlm\, \qty(\frac{r}{R})^{\delta_\ell^-})\star_r Y_\vlm~,              \\
		\eval{\qty(\star_{\scriptscriptstyle d} \f{\widetilde{J}}{d-1})}_{\Sigma_r} & = \frac{1}{\sqrt{2}} \sum_{\ell,\vec{m}} \sech(2 v_\ell)\qty(\ex{v_\ell} C_\vlm\, \qty(\frac{r}{R})^{\delta_\ell^+}-\ex{-v_\ell} B_\vlm\, \qty(\frac{r}{R})^{\delta_\ell^-})\star_r \widetilde{Y}_\vlm~.
	\end{aligned}
\end{equation}

In the above, \(Y_\vlm\) and \(\widetilde{Y}_\vlm\) are normalised scalar and \((d-2)\)-form spherical harmonics, respectively (see \cref{app:spherical harmonics} for conventions), \(\dd{\Omega_{d-1}}\) is the area element of a unit \((d-1)\)-sphere, and we have defined for convenience
\begin{equation}\label{eq:vl-def}
	v_\ell = \frac{1}{4}\log(\frac{\ell+d-2}{\ell})~,
\end{equation}
and the scaling exponents of the smooth and the divergent modes as
\begin{equation}\label{eq:scaling-exponents}
	\delta_\ell^\pm = -\frac{1}{2}\pm\qty(\ell + \frac{d-2}{2})~.
\end{equation}
Finally, the coefficients \(B_\vlm\) and \(C_\vlm\) are given as
\begin{equation}\label{eq:Blm-Qlms}
	\begin{aligned}
		B_\vlm & = \frac{1}{\sqrt{2}}\qty(\ex{v_\ell}Q_\vlm + \ii \ex{-v_\ell} \widetilde{Q}_\vlm) \qq{and} \\
		C_\vlm & = \frac{1}{\sqrt{2}}\qty(\ex{-v_\ell}Q_\vlm - \ii \ex{v_\ell} \widetilde{Q}_\vlm)~.
	\end{aligned}
\end{equation}

As a preliminary remark, note that in \(d=2\), the coefficients of the \emph{smooth} modes, \(\sim\qty(r/R)^{\delta_\ell^+}\) and \emph{divergent} modes, \(\sim\qty(r/R)^{\delta_\ell^-}\) correspond precisely to the \emph{creation} and \emph{annihilation} operators, \(A_\vlm\) and \(A_\vlm^\dagger\) defined in \cref{eq:ladder-def}. This identification breaks down in \(d>2\). From a technical perspective, this discrepancy is the precise origin of the Bogoliubov transformation anticipated by the intuitive argument presented above.

Inverting the above solution, yields expressions for \(Q_0\), \(B_\vlm\), and \(C_\vlm\) that can be inserted in the path integral. This is of course equivalent to solving for \(Q_\vlm\) or \(\widetilde{Q}_\vlm\), but it will be more convienient for our purposes. The zero mode takes the expected form:
\begin{equation}\label{eq:Q0}
	Q_0 = \frac{1}{\V^{1/2}\, R^{\frac{d-1}{2}}}\int_{\Sigma_r} \star_{\s{d}}\, \f{J}{1}~,
\end{equation}
while the higher modes are expressed as
\begin{align}\label{eq:Blm}
	B_{\ell\vm} = \qty(\frac{r}{R})^{-\delta_\ell^-}\frac{1}{\sqrt{2}}\int_{\Sigma_r}\qty(\ex{v_\ell} Y^*_{\ell\vm}\,\star_\s{d}\f{J}{1}-\ex{-v_\ell} \widetilde{Y}^*_{\ell\vm}\w\star_\s{d}\, \f{\widetilde{J}}{d-1})~, \\
	C_{\ell\vm} = \qty(\frac{r}{R})^{-\delta_\ell^+}\frac{1}{\sqrt{2}}\int_{\Sigma_r}\qty(\ex{-v_\ell} Y^*_{\ell\vm}\,\star_\s{d} \f{J}{1}+\ex{v_\ell} \widetilde{Y}^*_{\ell\vm}\w\star_\s{d}\, \f{\widetilde{J}}{d-1})~.  \label{eq:Clm}
\end{align}

Some comments are in order here. First, in the previous sections we postponed solving the constraint equation \cref{eq:constraint-free}. Equations \cref{eq:Blm,eq:Clm} provide the full solution. With it, alongside with the discussion in \cref{sec:symmetries}, it carries the weight of sealing the proof that the superfluid effective action has infinitely  many conserved charges. Second, coming back to the task at hand, upon inserting \cref{eq:Blm,eq:Clm} in the path integral, only the smooth parts of \(J\) and \(\widetilde{J}\) contribute. This follows directly by the fact that the path integral is weighted by the action, which nulls divergent contributions. Finally, henceforth we will concentrate on the modes \(B_{\vlm}\) as these are the relevant ones for preparing \emph{ket states}, obtained by path integrating from the centre of the ball, outwards to the boundary at \(r=R\). The complementary modes, \(C_{\vlm}\), play a role for preparing \emph{bra states}, constructed by an inward path integral from infinity to \(R\).\footnote{In this approach, bra states more naturally correspond to codimension-1 operators, rather than local operators.}

\subsubsection*{Matching the states}

Armed with \cref{eq:Q0,eq:Blm} we are ready for the final punch. First let us show that the states \(\ket{V_p}\) have fixed global charge. This is straightforward:
\begin{equation}\label{eq:Q0-Vp}
	Q_0\ket{V_p} = \frac{1}{\V^{1/2}\, R^{\frac{d-1}{2}}}\;\lim_{r\to 0} \int_{\B^d_R} \DD{\phi}\ex{-S[\phi]} \qty(\int_{\Sigma_r} \star_{\s{d}}\, \f{J}{1})\times V_p(0) = \frac{p}{\g\, \V^{1/2} R^{\frac{d-1}{2}}} \ket{V_p}~.
\end{equation}
In the above we used the OPE between the vertex operator and the current, in which the only component that contributes is
\begin{equation}\label{eq:OPE-J-V}
	J_r \times V_p(0) \sim \frac{\ii\,p}{r^{d-1}} V_p(0)~.
\end{equation}
The radial dependence is then subsequently cancelled by the volume form, \(r^{d-1} \dd{\Omega_{d-1}}\), in the integral over \(\Sigma_r\). \cref{eq:Q0-Vp} confirms the earlier claim: radial evolution does not mix the various fixed-charge superselection sectors --- it can at most mix states within each sector.

Such a mixing does indeed occur. It becomes evident upon observing that the states \(\ket{V_p}\) are annihilated by all \(B_\vlm\) modes. The computation is similar:
\begin{align}
	B_\vlm \ket{V_p} & = \lim_{r\to 0}\int_{\B^d_R} \DD{\phi}\ex{-S[\phi]}\; \cref{eq:Blm} \times V_p(0) \nn
	                 & \sim \lim_{r\to 0}\int_{\B^d_R} \DD{\phi}\ex{-S[\phi]}\; V_p(0) \qty(\frac{r}{R})^{2\ell + d-2} = 0~, \quad \t{for all \(\ell\) and \(\vec{m}\)}~, \label{eq:Blm-Vp}
\end{align}
where we used again the OPE \cref{eq:OPE-J-V}, the form \cref{eq:Blm} of the modes and that the only mode that survives scales as
\begin{equation}
	\qty(\frac{r}{R})^{\delta_\ell^+-\delta_\ell^-} = \qty(\frac{r}{R})^{2\ell+d-2}~,
\end{equation}
which vanishes in the limit \(r\to 0\).

The equations \cref{eq:Q0-Vp,eq:Blm-Vp} that \(\ket{V_p}\) satisfy bear a striking resemblance to \cref{eq:primary-conditions} that define \emph{primary} or highest-weight states. There is, however, one remaining challenge. \(B_\vlm\) is \emph{not} the annihilation operator \(A_\vlm\).  Nonetheless the two are related by a Bogoliubov transformation:\footnote{Here we have rescaled \(B_\vlm\) by a factor of \(\sqrt{\g}\).}
\begin{equation}\label{eq:Bogoliubov}
	B_\vlm = \cosh(v_\ell)\, A_\vlm + \sinh(v_\ell)\, A_\vlm^\dagger~.
\end{equation}
with \emph{squeeze parameter} \(v_\ell\) as in \cref{eq:vl-def}. It is clear that \(B_\vlm\) and its adjoint obey the same commutation relations as the original oscillators:
\begin{equation}
	\comm{B_\vlm^{\phantom{\dagger}}}{B^\dagger_{\ell'\vec{m}'}} = \frac{\sqrt{\ell(\ell+d-2)}}{R} \delta_{\ell,\ell'}\delta_{\vm,\vm'}~.
\end{equation}
As a result, the new oscillators furnish a different, yet equivalent, representation of the current algebra, and the states \(\ket{V_p}\) form the highest-weight states of this representation. Under this light, the vertex operators are \emph{primary operators}, not under the conformal algebra like their \(d=2\) counterparts, but under the Kac--Moody algebra that persists in higher dimensions.

A useful rewriting of \cref{eq:Bogoliubov} that is standard in quantum optics follows by introducing a \emph{squeezing operator},
\begin{equation}
	\cS_\vlm = \exp(-\frac{v_\ell}{2}\qty[A^2_\vlm-\qty(A^\dagger_\vlm)^2]) = \exp(-\frac{v_\ell}{2}\qty[B_\vlm^2-\qty(B^\dagger_\vlm)^2])~.
\end{equation}
In the terminology of quantum optics this is a one-mode squeezing operator. Our discussion will benefit by introducing an \emph{all-mode} squeezing operator:
\begin{equation}\label{eq:squeeeeze}
	\cS = \prod_{\ell,\vm} \cS_\vlm~.
\end{equation}
Let us spend a moment to discuss a couple of different interpretations of the squeezing operator. When expressed in terms of the Lorentzian oscillators, \(A_\vlm\),  \(\cS\) is a unitary operator that acts on the Hilbert space of \(\Sigma_R\), mapping energy eigenstates to \emph{squeezed states}. On the other hand, if expressed in terms the Euclidean oscillators, \(B_\vlm\), the squeezing operator can be viewed as a local operator, \(\cS(x)\), that can be inserted in the path integral.\footnote{In the absence of something to link with. If there is something to link with, one first performs the OPE and then shrinks \(\cS\).} This is most cleanly seen from \cref{eq:Blm} in the limit \(r\to 0\), which reduces to the form of the descendant operators in \cref{eq:descendants}. In terms of the original compact scalar, \(\phi\), it is an extremely irrelevant operator containing an infinite number of derivatives. Each mode \(\ell,\vm\) contributes to a specific order in derivatives, but the product in \cref{eq:squeeeeze} ensures that all modes are turned on. Nonetheless, \(\cS(x)\) plays an important role.

In terms of the squeezing operators, the two sets of oscillators are related via
\begin{equation}
	B_\vlm = \cS^\dagger A_\vlm\, \cS \qq{and} B_\vlm^\dagger = \cS^\dagger_{\vphantom{\vlm}} A_\vlm^\dagger\, \cS~.
\end{equation}
This rewriting makes the first part of the state-operator correspondence clear: \emph{Vertex operators correspond to squeezed primary states}. In equations:
\begin{equation}
	\ket{V_p} = \cS^\dagger \ket{p}~.
\end{equation}
It is clear that the above states have the right properties. Inverting the above equation, it follows that the primary states correspond to \emph{squeezed vertex operators}
\begin{equation}\label{eq:S-V}
	\no{\cS V_p}\!(x) = \lim_{r\to 0} \cS\qty[\Sigma_r]\cdot V_p(x)~.
\end{equation}
An important and perhaps counter-intuitive consequence of the above is that the lowest energy state, the vacuum \(\ket{0}\), is \emph{not} prepared by the empty path integral! Instead, it is prepared by the squeezing operator itself:
\begin{equation}
	\ket{0} \leftrightsquigarrow \cS(x)~.
\end{equation}
While peculiar, such a situation is not entirely novel. In three-dimensional CFTs, there is compelling evidence that on slices other than the sphere an empty path integral does \emph{not} prepare the ground state \cite{Belin:2018jtf}. In four-dimensional CFTs with generalised symmetries there exists a very similar statement \cite{Hofman:2024oze}: the vacuum on \(\S^1\times\S^2\) is prepared by a squeezing (there \emph{line}) operator. In both of the above cases it was the geometry --- particularly the lack of a conformal transformation from the Euclidean filling to the Lorentzian spacetime --- that was responsible for the excess of energy in the state produced by the empty path integral. In this case, it is the lack of the other standard ingredient of a state-operator correspondence in CFT: Weyl invariance. Nonetheless, the result is the same.

Having identified all the primary states the rest of the Hilbert space follows from representation theory. Excited states are constructed by dressing vertex operators by creation operators \(B_\vlm^\dagger\), expressed in terms of the currents. We will call such operators \emph{descendants}, in analogy with the states they create. At the last step, the descendants also get dressed with a squeezing operator.  To illustrate, excited states with one or two insertions are
\begin{align}
	A_\vlm^\dagger \ket{p}                       & \leftrightsquigarrow \no{\cS B_\vlm^\dagger V_p}\!(x)~,                       \\
	A_{\ell'\vm'}^\dagger A_\vlm^\dagger \ket{p} & \leftrightsquigarrow \no{\cS B_{\ell'\vm'}^\dagger B_\vlm^\dagger V_p}\!(x)~,
\end{align}
where the ordering is understood in the sense of \cref{eq:S-V}, in a nested way. The pattern continues and one can reach this way a generic state. This concludes the construction of the state-operator correspondence.

Lastly, let us briefly reemphasise the fact that inserting a single vertex operator in the path integral --- without dressing with the squeezing operator --- not only does not give a primary state, but not even an energy eigenstate, as measured by the cylinder Hamiltonian. Nonetheless, we can compute the average energy of a vertex operator state. Straightforwardly this gives
\begin{equation}
	\mel{V_p}{H}{V_p} = \Delta_p + \cE_0~,
\end{equation}
with
\begin{equation}
	\cE_0 = \frac{1}{R}\sum_{\ell,\vec{m}} \sinh[2](v_\ell) \sqrt{\lambda_\ell} = \frac{1}{R}\sum_{\ell=1}^\infty D_\ell \qty(\sqrt{\ell}-\sqrt{\ell+d-2})^2~,
\end{equation}
where
\begin{equation}
	D_\ell = \frac{(2\ell+d-2)\ (d+\ell-3)!}{\ell!\;(d-2)!}
\end{equation}
is the degeneracy of eigenvalues. \(\cE_0\) is essentially a zero-point energy, indicating that the cylinder Hamiltonian is not normal ordered with respect to the Euclidean oscillators. Do note that, as a consistency check, in \(d=2\) there is no zero-point shift.

\section{Superfluid phonons}\label{sec:superfluid}

A slight modification of the above analysis results in a state-operator correspondence for phonon excitations on a superfluid background. To get to that we return to the superfluid effective action, \cref{eq:superfluid-action}, which we repeat here for convenience:
\begin{equation}\label{eq:superfluid-action-copy}
	S[\xi] = \int \dd[d]{x} \p\qty(\partial_\mu \xi\, \partial^\mu \xi)~.
\end{equation}
For the time being we take spacetime to be flat, with metric \(\eta_{\mu\nu}\) of mostly plus signature. This choice is made purely for convenience --- to avoid polluting the forthcoming expressions with factors of the metric and its determinant. All results and constructions presented here admit a straightforward expression to general spacetimes.

It is standard practice to expand the field around a background configuration, \(\xi_0(x)\) as
\begin{equation}
	\xi(x) = \xi_0(x) + \phi(x)~,
\end{equation}
to study the dynamics of the phonons \(\phi(x)\). For a constant chemical potential \(\mu\), a typical background configuration is \(\xi_0(x)=\mu\,t\) (see e.g. \cite{Nicolis:2013lma}), or a generalisation thereof to a generic frame \cite{Kourkoulou:2022doz}. Recalling the discussion in \cref{sec:symmetries}, spontaneous symmetry breaking occurs for \(\mu\geq \mu_\t{crit}\) which depends on the specific microscopic realisation of the model. For our purposes we do not need to restrict to a specific background configuration, besides certain stability conditions for the phonon effective action, which we come to immediately.

Expanding to quadratic order, the effective action for the phonons takes the form
\begin{equation}
	S[\phi] = \frac{1}{2}\int \dd[d]{x} Z^{\mu\nu}\, \qty(\partial_{\mu} \phi+B_\mu)\qty(\partial_\nu \phi+B_\nu)~,
\end{equation}
where
\begin{align}
	Z^{\mu\nu} & = 2\p'_0\, \eta^{\mu\nu} +4\p''_0\, \pd^\mu \xi_0\, \pd^\nu \xi_0~, \qq{and} 	B_\mu = \frac{1}{2} \p'_0\, Z_{\mu\nu} \pd^\nu \xi_0~.
\end{align}
with \(\p'_0\) and \(\p''_0\) denoting the first and second derivatives of \(\p(\cdot)\) evaluated at the configuration \(\xi_0\). For general background, \(Z_{\mu\nu}\) is position dependent. Moreover, \(B_\mu\) couples to the phonons exactly like a gauge field for the \(\U(1)\) shift symmetry. Then one can turn on a background gauge field, \(\cA_\mu\), like in \cref{eq:covariant-der} that counterbalances \(B_\mu\). This is in fact common practice, see e.g. \cite{Son:2002zn,Delacretaz:2019brr}. Doing so allows us to turn \(B_\mu\) off in what follows. Note also that for a constant chemical potential --- to which we restrict henceforth --- \(\cA_\mu\) is necessarily flat.

Perturbative stability of the EFT is guaranteed \cite{Nicolis:2004qq,Kourkoulou:2022doz} if:
\begin{equation}\label{eq:stability-cond}
	Z^{00} > 0 \qq{and} Z^{0i}Z^{0j} - Z^{00} Z^{ij} \succ 0~.
\end{equation}
As a side-note, these conditions are compatible with sign-definiteness of \(\p'\), which was required in \cref{sec:symmetries}, whenever the background configuration, \(\xi_0\),  is globally timelike. This is the standard situation. See \cite{Kourkoulou:2023xqe} for an alternative situation and its stability analysis. Under these conditions, which we will henceforth assume, the phonon EFT can be treated as a bona fide theory.

\subsection{Symmetry considerations}

As per the general discussion in \cref{sec:symmetries}, this system also enjoys a \(\gf{\U(1)}{0}\times\gf{\U(1)}{d-2}\) symmetry. The \textquote{electric} 1-form current, corresponding to the zero-form symmetry now takes the form
\begin{align}\label{eq:eucl-current}
	J_\mu & = 2\p_0'\,\qty\big(\pd_\nu \phi + u_\mu\, u^\nu \pd_\nu \phi)~,
\end{align}
with
\begin{equation}\label{eq:u-def}
	u_\mu = \sqrt{\frac{2\p''_0}{\p'_0}} \pd_\mu \xi_0~.
\end{equation}
On the other hand, the \textquote{magnetic} \((d-1)\)-form current remains unmodified
\begin{equation}
	\f{\widetilde{J}}{d-1} = \star\dd{\phi}~,
\end{equation}
and is still conserved off-shell. Out of the two fundamental currents one can again construct dressed currents \(\cJ_f\) as defined in \cref{eq:dressed-current} and their associated conserved charges, \(\cQ_f[\Sigma]\). The dressing data consists of a scalar function, \(\f{f}{0}\), and a \((d-2)\)-form, \(\f{\widetilde{f}}{d-2}\), which are now required to satisfy:
\begin{alignat}{3}
	 & \qty(1+\abs{u}^2)\,\pd_\mu f+ (-1)^d u_\mu\, u^\nu \pd_\nu f + \frac{1}{2\p_0'}\varepsilon_{\mu}^{~\mu_2\cdots \mu_d}\pd_{\mu_2} \widetilde{f}_{\mu_3\cdots \mu_{d}} = 0~,
	\intertext{or more compactly in differential form language}
	 & \qty(1+\abs{u}^2) \star \dd{\f{f}{0}}+ (-1)^d\; u\w\iota_u\qty(\star\dd{\f{f}{0}}) + \frac{1}{2\p_0'}\dd{\f{\widetilde{f}}{d-2}} = 0~.
\end{alignat}
Here \(\abs{u} = u_\mu u^\mu\) and \(\iota_u\) denotes interior product. As discussed above, infinitely many solutions of these equations are guaranteed on general grounds providing a rich algebra of the conserved charges, \(\cQ_f\).

The algebra of these charges mirrors the structure of \cref{sec:compact-scalars}. Specifically, the commutator between two dressed charges takes the form:
\begin{align}
	\comm{\cQ_f}{\cQ_h} & = \ii\,(-1)^d\int_\Sigma \qty(\f{f}{0}\dd{\f{\widetilde{h}}{d-2}} - \f{h}{0}\dd{\f{\widetilde{f}}{d-2}})~.
\end{align}
Expanding the charges in a basis of modes (satisfying \cref{eq:beltramisation}) and applying the same method as in \cref{sec:compact-scalars}, one finds a familiar mode algebra for the non-zero modes:
\begin{equation}
	\comm{Q_n}{\widetilde{Q}_m} = \ii\sqrt{\lambda_n}\,\delta_{nm}~.
\end{equation}
This matches exactly the form of the Kac--Moody algebra in \cref{eq:KM-modes}.

\subsection{Canonical quantisation}

We now turn to the canonical quantisation of the phonon EFT on a \((d-1)\)-sphere of radius \(R\), \(\Sigma_R \equiv\S^{d-1}_R\). The task is again straightforward, as the theory is quadratic. Our aim here is quantise the theory in a way that lays bare the underlying symmetry structure.

The Hamiltonian follows immediately and reads \cite{Nicolis:2004qq}:
\begin{equation}
	H = \int_{\Sigma_R} \dd[d-1]{x}\sqrt{g_\Sigma}\ \frac{1}{2Z^{00}}\qty(\Pi_\phi^2 + K^{ij}\pd_i \phi \pd_j \phi)~.
\end{equation}
In the above \(\Pi_\phi\) is the momentum conjugate to \(\phi\) and \(K^{ij}\) reads
\begin{equation}
	K^{ij} = Z^{0i}Z^{0j} - Z^{00}Z^{ij}.
\end{equation}
The stability conditions discussed earlier ensure that the Hamiltonian is positive-definite. Relatedly, the matrix \(K^{ij}\) defines a positive-definite quadratic form on 1-forms over \(\Sigma_R\) given by:
\begin{equation}
	K\qty[\f{f}{1},\f{h}{1}] = \int_{\Sigma_R} \dd[d-1]{x}\sqrt{g_\Sigma}\   K^{ij} f_i h_j~,
\end{equation}
writing also \(K[f] \equiv K[f,f]\), for brevity. With this in hand, the Hamiltonian admits again a Sugawara-like expression in terms of the fundamental currents:
\begin{equation}
	H = \frac{\g}{2}\qty(\norm{\qty(\star_{\scriptscriptstyle d} \f{J}{1})_{\Sigma}}^2 + K\qty[\qty(\star_{\scriptscriptstyle d} \f{\widetilde{J}}{d-1})_{\Sigma}])~,
\end{equation}
where the subscript \(\Sigma\) denotes restriction on the spatial slice and \(\g=1/Z^{00}\).

This is nothing but a collection of coupled harmonic oscillators, as can be easily seen by expanding the currents in spherical harmonics as before. Each mode corresponds to a particular angular momentum label \((\ell,\vm)\), and the role of the spring matrix is played by the quadratic form. Explicitly:
\begin{align}
	H & = \frac{\g}{2}Q_0^2 + \sum_{\ell,\vm} \qty(Q_{\vlm}Q_{\vlm} + K^{\vlm,\ell'\vm'} \widetilde{Q}_\vlm\widetilde{Q}_{\ell'\vm'}) \\
	  & = \frac{\g}{2} Q_0^2 + \sum_{\ell,\vm} \kappa_{\vlm} A_\vlm^\dagger A_\vlm~,
\end{align}
with \(K^{\vlm,\ell'\vm'} = K\qty\big[\star \widetilde{Y}_{\vlm},\star \widetilde{Y}_{\ell'\vm'}]\). In the second line we have already diagonalised the Hamiltonian in terms of ladder operators \(A_\vlm\) and \(A_\vlm^\dagger\), satisfying, as before
\begin{equation}
	\comm{A_\vlm}{A_{\ell'\vm'}^\dagger} = \sqrt{\lambda_\ell} \delta_{\ell,\ell'} \delta_{\vm,\vm'}~,
\end{equation}
and \(\kappa_\vlm>0\) are the eigenvalues of the matrix \(K^{\vlm,\ell'\vm'}\).

It is immediately clear that everything that we derived above for the compact scalar extends for the superfluid phonons. The Hilbert space once again splits in superselection sectors graded by the global \(\U(1)\) charge:\footnote{Flux quantisation here, too, follows by a version of electric-magnetic duality.}
\begin{equation}
	\int_\Sigma \star_\s{d}\, \f{J}{1} = \frac{p}{\g}~, \qquad p\in\Z~.
\end{equation}
Furthermore the ladder operators endow each superselection sector with a Verma module structure, over the underlying Kac--Moody algebra. Altogether, the Hilbert space is consists of primary states \(\ket{p}\) satisfying
\begin{equation}
	Q_0 \ket{p} = \frac{p}{\g\;\V^{1/2}R^{\frac{d-1}{2}}} \ket{p} \qq{and}
	A_\vlm \ket{p} = 0 \quad \t{for all}\ \ell, \vec{m}~.
\end{equation}
And descendants built by acting with \(A_\vlm^\dagger\). What differs is only the energy levels, which are modified to include the eigenvalues of the spring matrix. For instance the state \(A_\vlm^\dagger\ket{p}\) has energy
\begin{equation}
	E_{A_\vlm^\dagger\ket{p}} = \Delta_p + \frac{\kappa_\vlm \sqrt{\ell(\ell+d-2)}}{R}~.
\end{equation}
with \(\Delta_p\) as in \cref{eq:primary-energy}.

\subsection{The other side}

On the operator side, the compactness of \(\phi\) still ensures that the only allowed operators are vertex operators labelled by integers:
\begin{equation}
	V_p(x) = {\ex{\ii p \phi(x)}}, \qquad p\in\Z~,
\end{equation}
along with the two currents, and composite, \emph{descendant} operators built from these ingredients. As before, the preferred basis to organise the descendants is the disorder one, \cref{eq:descendants}. What changes (slightly) is the OPE, now controlled by the two-point function:
\begin{equation}
	\ev{\phi(x) \phi(0)} = \qty(Z^{\t{E}}_{\mu\nu}\, x^\mu x^\nu)^{-\frac{d-2}{2}} + \t{gauge-dependent terms.}
\end{equation}
The gauge-dependent piece reflects the need to choose a representative of the equivalence class \(\phi\sim \phi+2\pi\). The matrix \(Z_\t{E}^{\mu\nu}\) is the Euclidean version of the kinetic matrix:
\begin{equation}
	Z_\t{E}^{00} = Z^{00}~, \qquad Z_\t{E}^{0i} = -\ii Z^{0i}~, \qq{and} Z_\t{E}^{ij} = - Z^{ij}~,
\end{equation}
and \(Z^{\t{E}}_{\mu\nu}\) is its inverse. Note that the stability conditions in \cref{eq:stability-cond} guarantee that \(Z_\t{E}^{\mu\nu}\) is positive-definite, so the two-point function is well-behaved.

States are built in the standard way by inserting an operator in the centre of the ball and performing the radial path-integral outwards. To compare to the states built canonically one must, once again, solve the radial evolution of the charges, or equivalently the conservation equations:
\begin{align}
	\dd\star_\s{d} \f{J}{1} & = 0 \qq{and}	\dd\star_\s{d} \f{\widetilde{J}}{d-1} = 0~,
\end{align}
on the \(d\)-dimensional Euclidean ball. What complicates matters is the rule by which the radial components get eliminated, since the two currents are no longer collinear. As follows from \cref{eq:eucl-current} the two currents are related by:
\begin{align}\label{eq:tweaked-duality}
	-\ii \f{J}{1} & = 2\p_0'\qty(\star_\s{d}\f{\widetilde{J}}{d-1}+ u \w\iota_u\qty(\star_\s{d} \f{\widetilde{J}}{d-1}))~,
	\intertext{or in components}
	-\ii J_\mu    & = 2\p_0'\qty( \qty(\star_\s{d}\widetilde{J}~)_{\!\mu} + u_\mu\ u^\nu\,\qty(\star_\s{d}\widetilde{J}~)_{\!\nu})~,
\end{align}
with \(u_\mu\) as in \cref{eq:u-def}. We actually do not need to solve the radial evolution explicitly. The results of \cref{sec:compact-scalars} and the argument around \cref{eq:constraint1} that non-linearity does not destroy any of the conserved dressed charges are sufficient to establish the result. This is so because the above equation is continuously connected to the free scalar by tuning \(\p''_0\), and so \(\abs{u}\), to zero.

Nonetheless, let us present a more convincing argument. For this argument we will pick the typical superfluid background, \(\xi_0=\mu\,t\), in the frame of the fluid. On the Euclidean ball, then, \(u_\mu\) is constant and pointing only in the radial direction. Hence the radial evolution equations become:\footnote{Here, since \(\p_0'\) is a constant we rescale \(\f{\widetilde{J}}{d-1}\) by \(2\p_0'\) to simplify the presentation.}
\begin{equation}\label{eq:RE-phonon}
	\begin{aligned}
		\pd_\r J_{\Sigma_\r} + \ii\, \qty(1+\abs{u}^2)\,\sd\star_\r\widetilde{J}_{\Sigma_\r} & = 0~, \\
		\pd_\r \widetilde{J}_{\Sigma_\r} + \ii\, \sd\star_\r J_{\Sigma_\r}                   & = 0~,
	\end{aligned}
\end{equation}
The solutions of this equation are identical to \cref{eq:currents-solved}. The only difference lies in the scaling exponents, which now take the form:
\begin{equation}
	\delta_\ell^\pm = -\frac{1}{2}\pm \sqrt{\ell(\ell+d-2)\qty(1+\abs{u}^2)+ \qty(\frac{d-2}{2})^2}~,
\end{equation}
and the squeezing parameter, now being:
\begin{equation}
	v_\ell = \frac{1}{2}\log(\frac{\sqrt{\ell(\ell+d-2)\qty(1+\abs{u}^2)+\qty(\frac{d-2}{2})^2}+\frac{d-2}{2}}{\sqrt{\ell(\ell+d-2)\qty(1+\abs{u}^2)+\qty(\frac{d-2}{2})^2}-\frac{d-2}{2}})~.
\end{equation}
The physics is clear. There is still one smooth and one divergent mode. The coefficient of the divergent mode, when expressed in terms of the currents becomes the Euclidean annihilation operator. It is related to the Lorentzian one, \(A_\vlm\) by a squeezing transformation  with parameter \(v_\ell\). Note that in this setup the spring matrix is proportional to the identity. The problem is entirely mapped to that of the previous section and therefore all its conclusions hold. Clearly, switching to a general frame does not spoil this result.

Just for illustration purposes, let us present how the scaling exponents and the squeezing parameter differ from the free case in a simple example of physical relevance, the conformal superfluid \cite{Hellerman:2015nra,Monin:2016jmo}.\footnote{Although the action is scale invariant, it is non-analytic at the origin. Expanding around a classical background introduces a scale, the chemical potential.}  In this setting, the equation of state dictates that \(\P = c\, \mu^d\), with \(c\) a dimensionless constant that cannot be fixed from first principles without invoking UV data. Working around the background configuration \(\xi_0 = \mu\,t\), one quickly finds
\begin{equation}
	\abs{u}^2 = d-2~.
\end{equation}
Interestingly, the value of \(u\) does not depend on the chemical potential.  The modified scaling exponents are then
\begin{equation}
	\delta_\ell^\pm = -\frac{1}{2}\pm \sqrt{\ell(\ell+d-2)\qty(d-1)+ \qty(\frac{d-2}{2})^2}~,
\end{equation}
while the squeezing parameter becomes
\begin{equation}
	v_\ell = \frac{1}{2}\log(\frac{\sqrt{\ell(\ell+d-2)\qty(d-1)+\qty(\frac{d-2}{2})^2}+\frac{d-2}{2}}{\sqrt{\ell(\ell+d-2)\qty(d-1)+\qty(\frac{d-2}{2})^2}-\frac{d-2}{2}})~.
\end{equation}

Returning to the general story, this section has established that the state–operator correspondence continues to hold for phonon excitations on a superfluid background. The overall features mirror those of the correspondence worked out in \cref{sec:compact-scalars} for the free scalar, with the necessary adjustments to account for the equation of state of the underlying superfluid.

\section{Conclusions and future directions}\label{sec:discussion}

In this paper we studied quantum field theories with emergent continuous higher-form symmetries with mixed anomalies. We focussed our interest in the case where the symmetry at long distances is \(\gf{\U(1)}{0}\times\gf{\U(1)}{d-2}\) and the mixed anomaly is captured by the inflow action
\begin{equation}
	S_\t{anomaly} = \frac{\ii}{2\pi}\int \f{\cB}{d-1}\w\dd\f{\cA}{1}~.
\end{equation}
The main achievement of the paper is a one-to-one correspondence between energy eigenstates on a spatial sphere and local operators. In free examples we gave an explicit construction of the energy eigenstates by performing a radial path integral with local operator insertions. Rather than relying on conformal symmetry, the organising principle here is a tower of conserved charges forming an abelian current algebra with central extension. These charges, defined on codimension-one hypersurfaces, generate an infinite-dimensional symmetry that acts nontrivially on both the Hilbert space of states and the space of local operators, structuring them into highest-weight modules of primaries and descendants. While the essential features of the correspondence --- namely the current algebra and the set of operators --- remain intact when irrelevant interactions are turned on, it remains a challenge to verify explicitly whether the correspondence itself extends to that regime.

Along the way, we uncovered several interesting facts about this class of theories, which are of their own merit. Notably, gapless phases with this structure of symmetry always describe a superfluid, for a suitable equation of state. This construction, which we termed “superfluidisation,” provides a universal framework for capturing the anomaly and the extended symmetry it enforces. This framework is also minimal: the magnetic current is realised in the simplest way possible, as a topological symmetry. These ideas resonate well with related approaches in \cite{Delacretaz:2019brr,Hinterbichler:2022agn,Hinterbichler:2024cxn}. In this sense, the superfluid EFT plays the role once held by free field realisations in 2d CFTs, now generalised to higher dimensions and without necessity of conformal invariance.

From a condensed matter perspective \cite{Wen:2018zux}, emergent anomalous higher-form symmetries are often indicative of topological order. While much of this discussion traditionally applies to gapped phases, there is growing interest in their gapless counterparts  \cite{Verresen:2019igf,Thorngren:2020wet,volovik2019topologicalsuperfluids,Wang:2021boi,Wen:2023otf,Huang:2023pyk,Antinucci:2024ltv}. In this context, the universality of the superfluid EFT in capturing the infrared physics of such anomalies may serve as a useful foothold in understanding the landscape of gapless topological phases.

Another major outcome of this work is the construction of an infinite number of conserved charges in the universal superfluid phase, satisfying a Kac--Moody algebra with central extension. When conservation laws proliferate to this extent, something remarkable often follows. A classic precedent is the enhancement of conformal symmetry to the full Virasoro algebra in two dimensions. In our case, the appearance of a Kac--Moody structure hints towards a different kind of structure: an underlying integrability in the superfluid EFT. In two dimensions this has been explicitly demonstrated \cite{Dodelson:2023uuu}. More broadly, the role of Kac--Moody algebras in integrable systems is well appreciated \cite{Reshetikhin,Sfetsos:2013wia,Lacroix:2023gig}, suggesting that the algebra uncovered here is not merely an organising tool but potentially a hint towards a new integrable arena.

We close by describing several possible future directions and open questions.

\paragraph{Large charge}

In recent years, a wealth of insight has emerged around CFTs with global symmetries, driven by the so-called large charge expansion (see e.g. \cite{Hellerman:2015nra,Monin:2016jmo,Banerjee:2017fcx,Jafferis:2017zna,Orlando:2019skh,Arias-Tamargo:2019xld,Gaume:2020bmp,Cuomo:2020rgt,Antipin:2022hfe,Cuomo:2022kio,Badel:2022fya} for a representative sample). The basic idea is this: take a CFT with, say, a \(\U(1)\) global symmetry, and focus on sectors of large fixed charge, \(Q\gg 1\). Placing the theory on a spatial sphere of radius \(R\), a scale hierarchy develops. The conserved charge induces a nonzero density, scaling as \(\rho\sim Q/R^{d-1}\), which sets an effective chemical potential, \(\mu\). In this regime the (approximate) CFT can be seen as a Wilsonian effective action at cutoff \(\Lambda\) \cite{Hellerman:2015nra}. In terms of energy scales, there is a separation:
\begin{equation}
	\frac{1}{R}\ll \Lambda\ll \mu~.
\end{equation}

The low-energy degrees of freedom are expected to be captured by a conformal superfluid, with effective action:
\begin{equation}
	S_\t{eff}[\xi] = \frac{\g}{2} \int \dd[d]{x} \qty\Big(\pd_\mu \xi\, \pd^\mu \xi)^{\frac{d}{2}}+\cdots~,
\end{equation}
expanded around a charged background \(\xi_0 = \mu\,t\) and the dots denote derivatively suppressed terms \cite{Monin:2016jmo}. This corresponds to a superfluid with pressure scaling as \(\mu^{d}\), and is typically assumed in the literature. But with the perspective offered in this work, where the superfluid EFT emerges as a universal description of anomalous symmetry phases, one might instead try to derive this behaviour. Specifically, for conformal theories with a \(\U(1)\) symmetry, one may hope to recover the emergent higher-form structure and the associated superfluid dynamics directly from conformal invariance and the assumed global symmetry. In two dimensions this is immediate \cite{Weinberg:2012cd} and is related to the free-field realisation. The expectation is that \emph{superfluidisation} provides the higher-dimensional analogue.

Regardless, assuming the conformal superfluid description in the large charge sector, it seems almost compulsory to investigate what constraints the resulting Kac--Moody algebra imposes on the CFT spectrum. Once the spectrum of operators is reorganised according to the current algebra, it is natural to ask whether operator dimensions, OPE coefficients, or selection rules are fixed or bounded by its structure. We leave such questions for future investigation.

\paragraph{Extended operators}

Another natural question is how extended operators fit into this picture. Recent developments in both quantum field theory and condensed matter have brought nonlocal operators to the forefront, driven largely by the growing importance of generalised global symmetries. Extended operators serve a role both as symmetry generators and the charged objects under higher-form symmetries. At the same time, there has been increasing interest in conformal defects, domain walls, and interfaces, especially in theories at or near criticality. Altogether, the message is clear: a complete understanding of quantum field theory must involve a deeper grasp of the physics of extended operators.

While the question of a state-operator correspondence for extended operators is subtle in general conformal field theories \cite{Belin:2018jtf}, a concrete proposal exists in four-dimensional CFTs with continuous higher-form symmetries: states on \(\S^2\times\S^1\)  correspond to line operators on \(\R^3\times\S^1\) \cite{Hofman:2024oze}. The methods employed there closely parallel those developed here and served as a key source of inspiration. It is natural to ask whether the present framework, combined with the technology of \cite{Hofman:2018lfz,Fliss:2023uiv}, can be extended to construct a state-operator map for extended operators in higher-form \cite{Armas:2018zbe,Delacretaz:2019brr,Armas:2023tyx} or even higher-group superfluids \cite{Armas:2024caa}. One potential complication in certain dimensions is the appearance of more relevant, Chern--Simons-like terms in the low-energy EFT \cite{Hinterbichler:2022agn}, driving the system to a gapped phase. This question is subject of immediate follow-up.

\paragraph{Non-abelian and non-invertible symmetries}

A natural question is how far these results extend to symmetry-breaking phases of non-abelian global symmetries. There is suggestive evidence pointing to the presence of analogous emergent symmetries and associated anomalies in such cases. When a continuous global symmetry \(G\) is spontaneously broken to a subgroup \(H\), the resulting Goldstone EFT, described by a \(G/H\) nonlinear sigma model, exhibits a rich structure of emergent higher-form symmetries \cite{Brauner:2020rtz}. See also \cite{Sheckler:2025rlk}.

When \(\pi_1(G/H)\) has infinite order, the theory admits a \((d-2)\)-form symmetry with current
\begin{equation}
	\f{\widetilde{J}}{d-1} = e_\fa \star \f{\omega}{1}^\fa,
\end{equation}
where \(\fa\) labels unbroken generators of \(G\), \(\f{\omega}{1}\) is the Maurer--Cartan form, and the coefficients \(e_\fa\) satisfy \(e_\fa f^\fa_{ij} = 0\), with \(f^k_{ij}\) the structure constants of \(G\). However, this structure is essentially captured by the \(\U(1)\) factors of \(G\), reducing the analysis to the abelian case already discussed. The anomaly \cref{eq:anomaly-action} also follows in this setting.

More generally, assuming \(G\) is simply connected and \(H\) connected, the effective theory exhibits a \(\gf{\U(1)}{d-3}\) symmetry for each \(\U(1)\) factor of the unbroken subgroup \(H\), following immediately by the topology of the underlying Lie groups. A particularly interesting case arises with a \(\gf{\U(1)}{d-4}\) symmetry, closely related to the Skyrmion number current \cite{Witten:1983ar}, which carries a mixed anomaly with the non-linearly realised part of the group \cite{Brauner:2020rtz}:
\begin{equation}
	\dd\star\f{\widetilde{J}}{d-3} = -\sum_{\sigma} c_\sigma \tr_\sigma\!\qty(\f{\cF}{2}\w\f{\cF}{2})~,
\end{equation}
with the sum over simple factors \(\sigma\) of \(G\). Several additional, typically composite, currents inherit anomalies from those above. It remains an open and intriguing question whether these elements can be unified into a broader framework and whether they yield an analogue of our construction for non-abelian SSB phases.

A different and rather compelling direction opens up when one considers symmetry-breaking phases involving continuous non-invertible symmetries \cite{Damia:2023gtc,GarciaEtxebarria:2022jky}. One simple yet rich example is obtained by gauging the discrete charge-conjugation symmetry \(\xi \to -\xi\) in the models discussed above. Once this is done, both \(\U(1)\) symmetries of the EFT are transmuted into non-invertible symmetries. These are generated by \textquote{cosine} topological operators:
\begin{align}
	D_\alpha[\Sigma_{d-1}] & = \cos\qty(\alpha \int_{\Sigma_{d-1}} \star \f{J}{1}) \qq{and} \widetilde{D}_\alpha[\Sigma_{1}] = \cos\qty(\alpha \int_{\Sigma_{1}} \star \f{\widetilde{J}}{d-1})~.
\end{align}
The parameter \(\alpha\) now takes values in the open interval \((0,\pi)\). At the endpoints \(\alpha = 0\) and \(\pi\) the defects remain invertible, but in between lies a continuum of non-invertible topological operators. As a result, the moduli space of vacua becomes an orbifold \cite{Damia:2023gtc}, with fixed points precisely at these endpoints. See also \cite{Chang:2020imq,Nguyen:2021yld,Heidenreich:2021xpr,Thorngren:2021yso,Bhardwaj:2022yxj,Antinucci:2022eat} for in-depth analysis of these non-invertible symmetries. Intriguingly, there exists a non-invertible analogue of the Kac--Moody algebra \cref{eq:comm-charges} on the gauged side \cite{Hofman:2024oze}, built from the dressed currents \cref{eq:dressed-current}:
\begin{equation}
	\cD_f\qty[\Sigma_{d-1}] = \cos(\int_{\Sigma_{d-1}} \star \cJ_f)~.
\end{equation}
Exploring the representation theory of this extended algebra might open the door to a non-invertible state-operator correspondence.

\paragraph{Entanglement entropy}

A fruitful application of the state-operator correspondence in CFTs is the calculation of entanglement entropy. The idea is as follows. Consider a pure state \(\rho=\ketbra{\psi}\) defined on a spatial sphere \(\S^{d-1}\) and let \(A\) be a chosen subregion. The entanglement entropy of this state, on \(A\)  is given by
\begin{equation}\label{eq:ee}
	\cS[\rho_A] = - \tr_{A}(\rho_A \log \rho_A)~.
\end{equation}
Here, the Hilbert space is assumed to factorise as \(\cH_\Sigma = \cH_A\otimes\cH_{\bar{A}}\). While this assumption is subtle and generally requires care --- see for instance \cite{Ohmori:2014eia} --- it will not affect the discussion at hand. The reduced density matrix \(\rho_A\) is defined by tracing out the complement: \(\rho_A = \tr_{{\bar{A}}}\rho\). A standard approach to evaluating \(\cS[\rho_A]\) is the replica trick. One first computes the Rényi entropies,
\begin{equation}
	\cS_n[\rho_A] = \frac{1}{1-n} \log \tr_A(\rho_A^n), \qquad \cS[\rho_A] = \lim_{n \to 1} \cS_n[\rho_A]~,
\end{equation}
analytically continuing to \(n\to 1\) to obtain the entanglement entropy. Using the state-operator correspondence, the state \(\ket{\psi}\) can be associated with a local operator \(\Psi\). Then \(\rho_A\) corresponds to a path integral with insertions of \(\Psi\)  and \(\Psi^\dagger\) on a geometry with a cut along \(A\), and the replicated trace becomes a path integral on the \(n\)-fold branched cover:
\begin{equation}
	\tr_{A} \rho_A^n = \frac{\ev{\Psi^\dagger(x_1)\cdots\Psi^\dagger(x_n)\Psi(y_1)\cdots\Psi(y_n)}_{\Sigma_n}}{\ev{\Psi^\dagger(x_1)\Psi(y_1)}_{\Sigma_1}^{n\vphantom{{}^n}}}~.
\end{equation}
Here, \(\Sigma_n\) is the so-called replica manifold; an \(n\)-fold branched cover of \(\R^d\), with the branch cut lying on \(A\).

For non-conformal theories, computing entanglement entropy is considerably more subtle, even in free cases. While vacuum entanglement is fairly well understood (see e.g. \cite{Casini:2009sr} for a review of massive free theories, and \cite{Agon:2013iva,Huerta:2022cqw} for results on abelian Goldstone bosons in three dimensions), the story for excited states remains largely uncharted. The state-operator correspondence developed here offers a concrete framework for addressing this gap, mirroring techniques familiar from CFTs. We expect this to offer a practical computational tool and to be of immediate relevance and interest.

\paragraph{Lessons for holography?}

An intriguing direction is whether the framework developed here can serve as a toy model for probing the construction or counting of states in holography. A seed of this idea appears in \cite{Benini:2022hzx}, where Chern--Simons theory with gauged one-form symmetries is proposed as a topological toy model for gravity, “holographically dual” to a WZW model. This picture was refined and extended in \cite{Antinucci:2024bcm}, which demonstrated that symmetry topological field theories (SymTFTs) with gauged higher-form symmetries can act as holographic duals to symmetry-breaking EFTs. The present work complements this proposal by providing an explicit state-operator correspondence in such EFTs --- at least in the abelian case. Understanding how these states are encoded in the bulk theory, in this controlled envoironment, could offer insight on more physically grounded holographic constructions.

\section*{Acknowledgements}
I thank Jeremías Aguilera-Damia, Riccardo Argurio, Jackson Fliss, Giovanni Galati, Eduardo García-Valdecasas, Ben Hoare, Diego Hofman, and Grégoire Mathys for helpful discussions, and especially Riccardo Argurio, Diego Hofman, and Grégoire Mathys for comments on an earlier draft. I would also like to credit PSI 2025 for hospitality during the final stages of this work and its participants for inspiring discussions. This work was supported by a Marina Solvay fellowship and by the Fonds de la Recherche Scientifique (FNRS) under grant \textnumero\ 4.4503.15.

\appendix

\section{Transversal spherical harmonics}\label{app:spherical harmonics}

For convenience, we collect here some key facts about \emph{scalar} and \emph{transversal \((d-2)\)-form} spherical harmonics in \((d-1)\)-dimensions. Spherical harmonics on a \((d-1)\)-sphere are labelled by \((d-1)\) integers, \(\ell\geq 0\) and \(\vec{m}=(m_1,\cdots,m_{d-2})\), satisfying the standard hierarchy:
\begin{equation}\label{eq:ell-ineq}
	\ell \geq m_1 \geq \cdots \geq m_{d-3}\geq \abs{m_{d-2}}~.
\end{equation}

On a \((d-1)\)-sphere of radius \(r\) the spherical harmonics, \(Y_\vlm\), are eigenfunctions of the Laplacian (defined with non-negative spectrum):
\begin{align}
	\lapl\, Y_\vlm & = \underset{\lambda_\ell(r)}{\underbrace{\frac{\ell(\ell+d-2)}{r^2}}}\; Y_\vlm~.
\end{align}
The explicit form of the harmonics in terms of angular coordinates is standard (see e.g. \cite{Higuchi:1986wu}) and not needed here. What matters is their orthonormality:
\begin{equation}
	\ip{Y_\vlm}{Y_{\ell\vm'}} = \int_{\S^{d-1}_r} Y^*_\vlm\w\star_{r} Y_{\ell'\vm'} = \int_{\S^{d-1}_r} \dd[d-1]x \sqrt{g}\; Y^*_\vlm\;Y_{\ell'\vm'} = \delta_{\ell,\ell'}\delta_{\vm,\vm'}~.
\end{equation}
and their degeneracy, which counts the number of independent harmonics with eigenvalue \(\lambda_\ell\):
\begin{equation}
	D_\ell = \frac{(2\ell+d-2)\ (d+\ell-3)!}{\ell!\;(d-2)!}~.
\end{equation}

To make the radius dependence explicit, we choose a normalisation such that
\begin{align}
	Y_\vlm & = r^{-\frac{d-1}{2}} Y^{(1)}_\vlm~,
\end{align}
where \(Y^{(1)}_\vlm\) are the harmonics on a unit sphere. In particular, the mode \(\ell=0\), the unique zero mode, is constant and normalised as:
\begin{equation}
	Y_{0} = \vol(\S^{d-1}_r)^{-\frac{1}{2}}~.
\end{equation}
Altogether \(\set{Y_\vlm}\) provide a complete orthonormal basis of functions on \(\S^{d-1}_r\).

Alongside scalar harmonics, in the main text we make heavy use of \emph{transversal \((d-2)\)-form} spherical harmonics.\footnote{Here we only discuss the case \(d>2\). The case \(d=2\) falls back to scalar trigonometric functions.} These generalise the familiar \textquote{magnetic} vector harmonics on \(\S^2\), such as \(\hat{\vec{n}}\times\nabla Y_{\ell m}\). On \(\S^{d-1}_r\) they take the form:
\begin{equation}\label{eq:transpherical}
	\widetilde{Y}_\vlm = \frac{(-1)^d}{\sqrt{\lambda_\ell(r)}} \star_r\!\dd{Y_\vlm} = \frac{(-1)^d\,r^{\frac{d-3}{2}}}{\ell(\ell+d-2)}\ \pd_i Y^{(1)}_\vlm\ \star_\s{1}\!\dd{x^i}~,
\end{equation}
where \(\star_\s{1}\) denotes the Hodge-star operator on a unit \((d-1)\)-sphere. These forms are manifestly transversal (coclosed):
\begin{equation}
	\cdd{\widetilde{Y}_\vlm}=0~,
\end{equation}
and satisfy the same Laplacian eigenvalue equation as their scalar parents
\begin{equation}
	\lapl\,\widetilde{Y}_\vlm = \lambda_\ell(r)\,\widetilde{Y}_\vlm~.
\end{equation}
They carry the same eigenvalues and degeneracies. Here \(\lapl = \qty\big(\dd+\cdd)^2\) is the Hodge--de Rham Laplacian on \((d-2)\)-forms. The prefactor \(\lambda_\ell(r)^{-1/2}\) in \cref{eq:transpherical} ensures orthonormality under the natural inner product:
\begin{equation}
	\ip{\widetilde{Y}_\vlm}{\widetilde{Y}_{\ell'\vm'}} = \int_{\S^{d-1}_r} \widetilde{Y}^*_\vlm\w\star_r \widetilde{Y}_{\ell'\vm'} = \delta_{\ell,\ell'}\delta_{\vm,\vm'}~,
\end{equation}
while the factor \((-1)^d\) is simply for convenience in some formulas in the main text. One detail is worth emphasising: there are no zero-modes. Since \(\widetilde{Y}_\vlm\) is build from a derivative of \(Y_\vlm\), the \(\ell=0\) mode, drops out. Normally one has to solve separately for the zero modes. Here, the triviality of the cohomology group \(\H^{d-2}(\S^{d-1}_r)\) in \(d>2\) guarantees the absence of zero-modes. Altogether, \(\set{\widetilde{Y}_\vlm}\) provide a complete orthonormal basis of transversal \((d-2)\)-forms on \(\S^{d-1}_r\).

\section{Details on the radial evolution}\label{app:radial-evol}

In this appendix, we elaborate on the radial evolution discussed in \cref{sec:compact-scalars}. Our goal is to solve \cref{eq:RE}, repeated below for convenience:
\begin{equation}\label{eq:RE-app}
	\begin{aligned}
		\pd_\r J_{\Sigma_\r} + \ii\, \sd\star_\r\widetilde{J}_{\Sigma_\r}  & = 0~, \\
		\pd_\r \widetilde{J}_{\Sigma_\r} + \ii\, \sd\star_\r J_{\Sigma_\r} & = 0~.
	\end{aligned}
\end{equation}
We begin by expanding the currents in spherical harmonics, \(Y_\vlm\) and \(\widetilde{Y}_\vlm\), introduced in \cref{app:spherical harmonics}:
\begin{equation}\label{eq:mode-exp-sphere}
	\begin{aligned}
		\eval{\qty(\star \f{\widetilde{J}}{d-1})}_{\Sigma_r} \equiv J_{\Sigma_\r}  & = \sum_{\ell,\vec{m}} \widetilde{J}_\vlm(\r)\,\star_\r \widetilde{Y}_\vlm(\r) ~,     \\
		\eval{\qty(\ii \star \f{J}{1})}_{\Sigma_r}\equiv \widetilde{J}_{\Sigma_\r} & = J_0(\r)\,\star_\r Y_{0}(r) + \sum_{\ell,\vec{m}} J_\vlm(\r)\,\star_\r Y_\vlm(\r)~.
	\end{aligned}
\end{equation}
This decomposition yields a system of ordinary differential equations (ODEs) for the modes. The expansion is such that the initial condition at \(\r=R\) is
\begin{equation}\label{eq:bc}
	J_0(R) = Q_0~, \qquad J_\vlm(R) = Q_\vlm~, \qq{and} \widetilde{J}_\vlm(R) = \widetilde{Q}_\vlm~.
\end{equation}

The resulting system of ODEs is:
\begin{align}\label{eq:RE0}
	\dv{r}J_0(\r) + \frac{d-1}{2\,\r} J_0(\r)                                                                                 & = 0                          \\
	\dv{\r}J_\vlm(\r) + \frac{d-1}{2\,\r} J_\vlm(\r) + \ii \frac{\sqrt{\ell(\ell+d-2)}}{\r} \widetilde{J}_\vlm(\r)            & = 0~,         \label{eq:REn} \\
	\dv{\r}\widetilde{J}_\vlm(\r) - \frac{d-3}{2\,\r} \widetilde{J}_\vlm(\r) - \ii \frac{\sqrt{\ell(\ell+d-2)}}{\r}J_\vlm(\r) & = 0~. \label{eq:REtn}
\end{align}
This system, together with the boundary conditions \cref{eq:bc}, admits a unique solution:
\begin{equation}
	J_0(\r) = Q_0 \qty(\frac{r}{R})^{-\frac{d-1}{2}}~,\label{eq:J0-soln}
\end{equation}
and
\begin{alignat}{5}
	J_{\vlm}(\r)             & =  \frac{\sech(2 v_\ell)}{\sqrt{2}}\Bigg(\ex{-v_\ell} &  & C_\vlm\, \qty(\frac{r}{R})^{\ell + \frac{d-3}{2}}+\ex{v_\ell}  &  & B_\vlm\, \qty(\frac{r}{R})^{-\ell - \frac{d-1}{2}}\Bigg) \label{eq:Jn-soln}    \\
	\widetilde{J}_{\vlm}(\r) & = \frac{\sech(2 v_\ell)}{\sqrt{2}}\Bigg(\ex{v_\ell}   &  & C_\vlm\, \qty(\frac{r}{R})^{\ell + \frac{d-3}{2}}-\ex{-v_\ell} &  & B_\vlm\, \qty(\frac{r}{R})^{-\ell - \frac{d-1}{2}}\Bigg)~, \label{eq:Jtn-soln}
\end{alignat}
with coefficients
\begin{equation}
	B_{\vlm} = \frac{1}{\sqrt{2}}\qty(\ex{v_\ell}Q_\vlm + \ii \ex{-v_\ell} \widetilde{Q}_\vlm) \qq{and} C_{\vlm} = \frac{1}{\sqrt{2}}\qty(\ex{-v_\ell}Q_\vlm - \ii \ex{v_\ell} \widetilde{Q}_\vlm)~.
\end{equation}
We also introduced for convenience a parameter \(v_\ell\) defined as:
\begin{equation}
	v_\ell = \frac{1}{4}\log(\frac{\ell+d-2}{\ell})~.
\end{equation}
Each mode features two components: one that remains smooth as \(r/R\to 0\), and one that diverges in that limit. This behaviour holds for all \(\ell\geq 1\). The corresponding scaling exponents governing these behaviours can be neatly repackaged as
\begin{equation}
	\delta_\ell^\pm = -\frac{1}{2}\pm \qty(\ell + \frac{d-2}{2})~.
\end{equation}

To connect with the standard approach in CFTs, where one seeks eigenfunctions of the dilatation operator, \(\cD = r\, \pd_r\), we can rewrite the radial evolution equations \cref{eq:RE0,eq:REn,eq:REtn} as
\begin{align}
	\cD J_0(r) + \frac{d-1}{2} J_0(r)                     & = 0~,       \\
	\cD \vec{V}_{\vlm}(r) + \bbD_{\ell} \vec{V}_{\vlm}(r) & = \vec{0}~,
\end{align}
where
\begin{equation}
	\vec{V}_{\vlm}(r) = \mqty(J_\vlm(r) \\[0.4em] \widetilde{J}_\vlm(r)) \qq{and} \bbD_{\ell} = \mqty(\frac{d-1}{2} && i \sqrt{\ell(\ell+d-2)} \\[0.4em] -i \sqrt{\ell(\ell+d-2)} && -\frac{d-3}{2})~.
\end{equation}
Diagonalising the above system yields immediately \cref{eq:J0-soln,eq:Jn-soln,eq:Jtn-soln}.

\printbibliography
\end{document}